\pdfoutput=1
\documentclass[11pt]{article}

\usepackage[final]{acl}

\usepackage{times}
\usepackage{latexsym}

\usepackage[T1]{fontenc}

\usepackage[utf8]{inputenc}

\usepackage{microtype}

\usepackage{inconsolata}

\usepackage{graphicx}

\usepackage{xspace}
\usepackage{graphicx}
\usepackage{stackengine}
\usepackage{verbatim}
\usepackage{caption}
\usepackage{subcaption}
\usepackage{listings}
\usepackage[noend]{algpseudocode}
\usepackage{framed}
\usepackage[skins]{tcolorbox}
\usepackage{amsmath}
\usepackage{algorithm}
\usepackage{algpseudocode}
\usepackage{multirow} 
\algrenewcommand\algorithmicindent{1.2em}
\usepackage{footmisc}

\newcommand{\tool}{\textsc{LeanCode}\xspace}
\newcommand{\toolbase}{\textsc{DietCode}\xspace}
\newcommand{\toolbasebase}{\textsc{SlimCode}\xspace}
\definecolor{custom-blue}{rgb}{0,0,0}
\newcommand{\code}[1]{{\footnotesize\texttt{#1}}}
\newcommand{\intuition}[1]{
\begin{tcolorbox}[colback=white,boxrule=1pt,top=0pt,bottom=0pt,left=1pt,right=2pt,top=2pt,bottom=2pt]
\em #1
\end{tcolorbox}
}

\title{{\tool}: Understanding Models Better for Code Simplification of Pre-trained Large Language Models}

\author{
  \textbf{Yan Wang\textsuperscript{1}\thanks{\ \ Equal contribution.}},
  \textbf{Ling Ding\textsuperscript{1}$^{*}$},
  \textbf{Tien N. Nguyen\textsuperscript{2}$^{*}$},
  \textbf{Shaohua Wang\textsuperscript{1}\thanks{\ \ Corresponding authors.}},
  \textbf{Yanan Zheng\textsuperscript{3}$^\dagger$}
  \\[1ex]
  \textsuperscript{1}Central University of Finance and Economics, \\
  \textsuperscript{2}University of Texas at Dallas,
  \textsuperscript{3}Yale University
  \\[1ex]
  {%
    {\{dayanking,imlingding\}@gmail.com},\;
    {tien.n.nguyen@utdallas.edu},\;
  }\\
  {%
    {davidshwang@ieee.org},\;
    {yanan.zheng@yale.edu}%
  }
}

\begin{document}
\maketitle
\begin{abstract}
Large Language Models for code often entail significant computational complexity, which grows significantly with the length of the input code sequence. 
We propose {\tool} for code simplification to reduce training and prediction time, leveraging {\em code contexts in utilizing attention scores} to represent the tokens' importance. We advocate for the selective removal of tokens based on the average {\em context-aware} attention scores rather than average scores across all inputs. {\tool} uses the attention scores of `CLS' tokens within the encoder for classification tasks, such as code search. It also employs the encoder-decoder attention scores to determine token significance for sequence-to-sequence tasks like code summarization.
Our evaluation shows {\tool}'s superiority over the SOTAs \toolbase and~{\toolbasebase}, with improvements of 60\% and 16\% for code search, and 29\% and 27\% for code summarization, respectively.
\end{abstract}

\vspace{-4pt}
\section{Introduction}
\label{sec:intro}

Pre-trained Large Language Models (PLLMs) demand significant computational resources, often constraining input word or code token lengths. For example, when using CodeBERT~\cite{codebert-emnlp20} locally, there's a limitation of 512 tokens. CodeT5~\cite{wang-etal-2021-codet5}, CodeGen~\cite{codegen}, and GPT-4~\cite{ChatGPT}~also entail high computational overheads and costs, particularly with longer input code sequences. Code simplification of PLLMs is a practical way to reduce training and prediction time, while maintaining performance of a PLLM as much as possible. Given various pre-trained models and downstream tasks, it is intuitive that not all input tokens play critical roles in downstream-tasks. To tackle this challenge, the state-of-the art approaches, like DietCode~\cite{dietcode-fse22} and SlimCode~\cite{wang2024natural}, were proposed to simply the input program of a PLLM.

First, DietCode~\cite{dietcode-fse22} 
computes the average self-attention score for each code token across various contexts(global attention scores), treating it as the representative importance score for the token across the entire dataset. It then employs this score to determine whether to eliminate the code token from all inputs. However, due to the inherent nature of a PLLM, the same code token may be associated with different self-attention weights across different contexts. In our experiments, we observed a wide range of self-attention weights assigned to the same code token depending on {\em the contexts in which it appears}. Consequently, assigning {\em an average score} to each token for its importance is not appropriate. 

Second, the way of DietCode using attention scores cannot directly reflect the importance of tokens in downstream tasks. 
Each input is represented by a special token labeled `CLS'. This `CLS' token's vector is computed based on the vectors of the constituent code tokens within that code snippet and description, determined by their self-attention scores. The vector of the `CLS' token is fed into the final fully-connected layer for classification after the encoder. Consequently, tokens with higher CLS-attention scores hold greater importance for the classification task compared to individual token self-attention scores. However, DietCode assesses the self-attention scores of all input tokens for classification purposes, lacking the focus on the `CLS' token, whose vector essentially encapsulates all necessary information for the classification.

Third, similarly, for code summarization, a transformer-based encoder-decoder architecture translates a given code snippet into a description. The self-attention mechanism within the encoder enables each token in the input to interact with all other tokens, effectively capturing  dependencies and contextual information across different positions within the sequence, particularly adept at capturing long-range dependencies. DietCode only uses the attention scores of this encoder to signify the importance of code tokens and discarded these encoder-decoder attention scores.

Alternatively, \toolbasebase~\cite{wang2024natural} is the state-of-the-art method based on human knowledge and uses a set of rules to determine the importance for different input code tokens. Specifically, it categorizes tokens into 8 priority levels based on the nature of the code tokens.
For example, tokens in method signatures receive the highest priority, while symbol tokens (e.g., brackets, separators, and operators) have the lowest. 
However, performing better than DietCode, \toolbasebase still has the following issues~\cite{wang2024natural}. The manually selected priority levels with only 8 tiers result in a large number of tokens having the same priority, thus making token removal lack a solid basis. Secondly, the simplified code is fed into the model to complete downstream tasks, but model cognition can differ from human cognition. Thus, tokens considered important by human knowledge may not necessarily be considered important in model knowledge, leading to unexpected results.

In this paper, we posit that 1) model knowledge is more suitable for code simplification. 2) when leveraging attention scores to indicate the importance of code tokens in code simplification, one needs to consider their appearance {\bf contexts}, the {\bf CLS} and {\bf encoder-decoder attention} are closely tied to downstream tasks, making their scores more appropriate for determining token importance in such tasks. 
We introduce a novel approach to code simplification, named {\tool}. {\tool}'s
overarching goal is to streamline computation {\color{black}{time}}, including those for training and inference, in downstream tasks like code search and summarization, while preserving performance to the fullest extent possible.
First, we advocate for the removal of a specific instance of token based on the attention score unique to that context of that occurrence, rather than relying on average scores across all inputs. We use the {\em statement type to represent the context} of that occurrence of $t$.
Second, we propose integrating the {\color{black}{self-}}attention scores of `CLS' tokens for classification tasks, such as code search. Finally, for sequence-to-sequence tasks, we consider encoder-decoder attention scores to ascertain the significance of input tokens. 

The contributions of this paper are as follows:


1. We carried out a systematic analysis of the significant tokens learned by both `CLS' and encoder-decoder attentions, comparing them with the tokens learned only by encoder self-attention.
    
2. We present a new context-aware, code simplification, which is used in discriminative and sequence-to-sequence {\color{black}{generation}} tasks. 

3. We evaluate {\tool} in two downstream tasks and the results show its superiority over \toolbase and \toolbasebase, with improvements of up to 60\% and 16\% for code search, and 29\% and 27\% for code summarization, respectively.


4. We evaluate LeanCode's cross-model transfer capability by feeding simplified code, generated using CodeT5, into the GPT-4o model to assess downstream task performance. 



\section{Preliminary Empirical Study}
\label{sec:empirical-study}


We conducted an empirical study on code search and summarization tasks to further investigate the significant tokens identified by the CLS and the Encoder-Decoder attentions.  
Our main focus is on the token level. The importance of statements and functions can be represented using tokens. We utilize the same models and datasets as \toolbase and \toolbasebase. 
Specifically, for the code-search classification task, we use CodeBERT (an encoder-only model) and the encoder of CodeT5 (an encoder–decoder model), each augmented with an additional fully connected layer. For code summarization, we employ CodeBERT with a Transformer decoder as well as the complete CodeT5 model. For all experiments, we measure token importance using CLS (or Encoder-Decoder) attention scores from the final layer of the encoders of CodeBERT and CodeT5 (or the decoders of transformer and CodeT5), since both models compute scores across multiple layers. Token weights from the final layer provide the accurate representations of contextual relationships.
Both tasks utilize the CodeSearchNet Corpus~\cite{husain2019codesearchnet}.

\begin{figure*}
 \centering
 \begin{subfigure}[h]{0.49\textwidth}
     \centering
     \includegraphics[width=\textwidth]{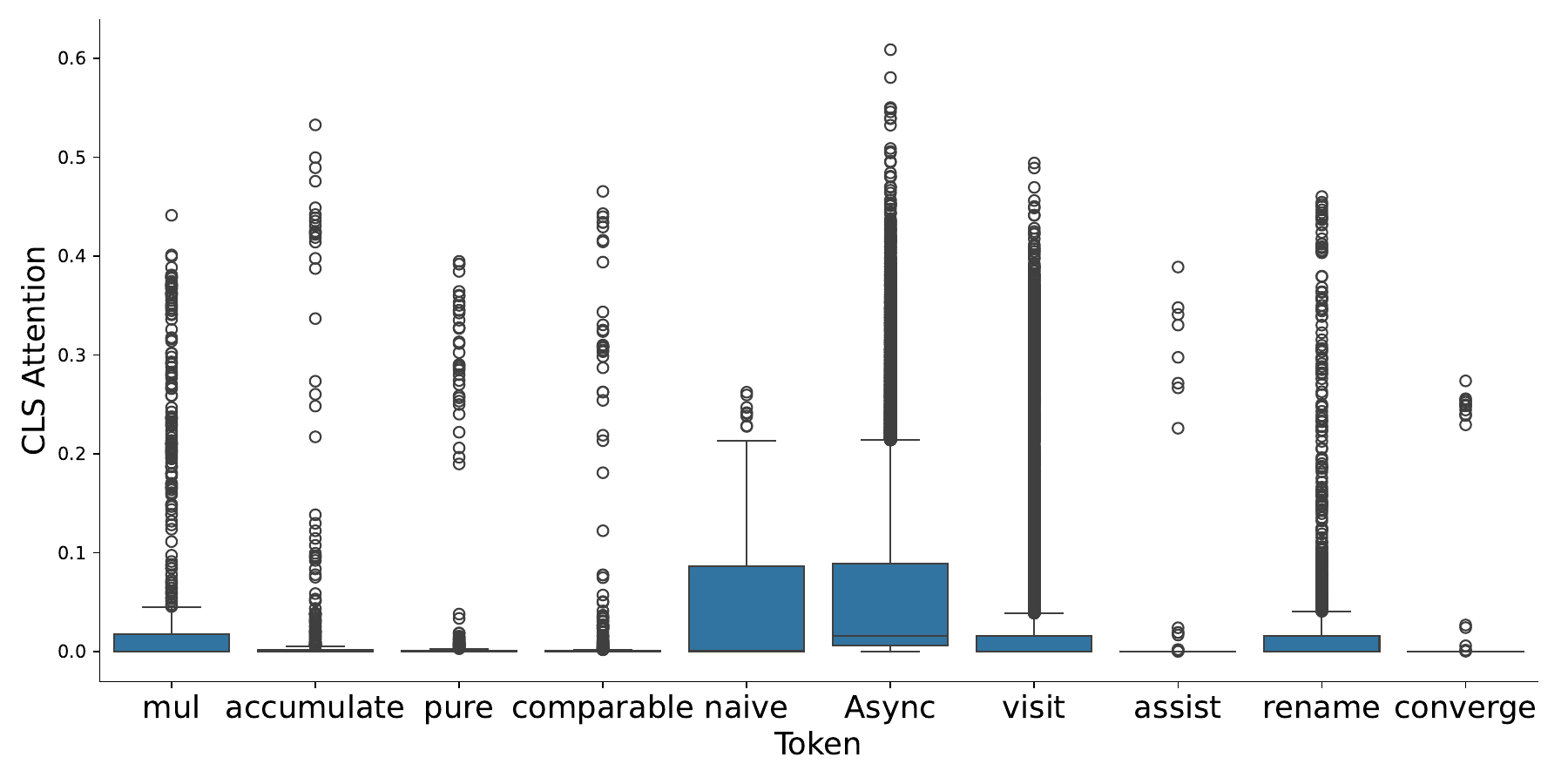}
     \vspace{-16pt}
     \caption{Top 10 tokens with highest variance in 
     CLS attention for code search.}
     \label{fig:cls-var-hi}
 \end{subfigure}
 \hfill
 \begin{subfigure}[h]{0.49\textwidth}
     \centering
     \includegraphics[width=\textwidth]{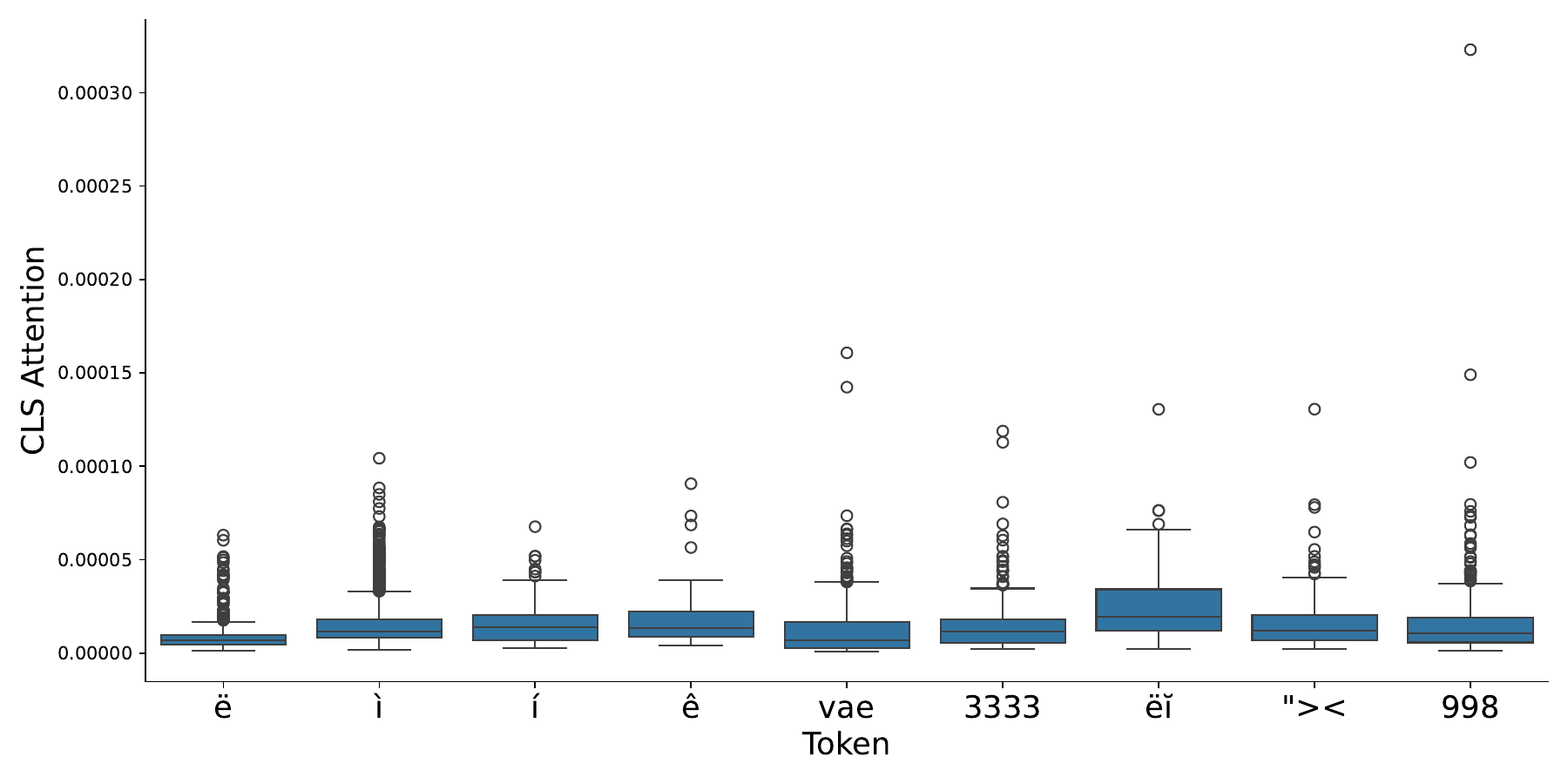}
     \vspace{-16pt}
     \caption{Bottom 10 tokens with lowest variance in CLS attention for code search.}
     \label{fig:cls-var-lo}
 \end{subfigure}
 \hfill
 \begin{subfigure}[h]{0.49\textwidth}
     \centering
     \includegraphics[width=\textwidth]{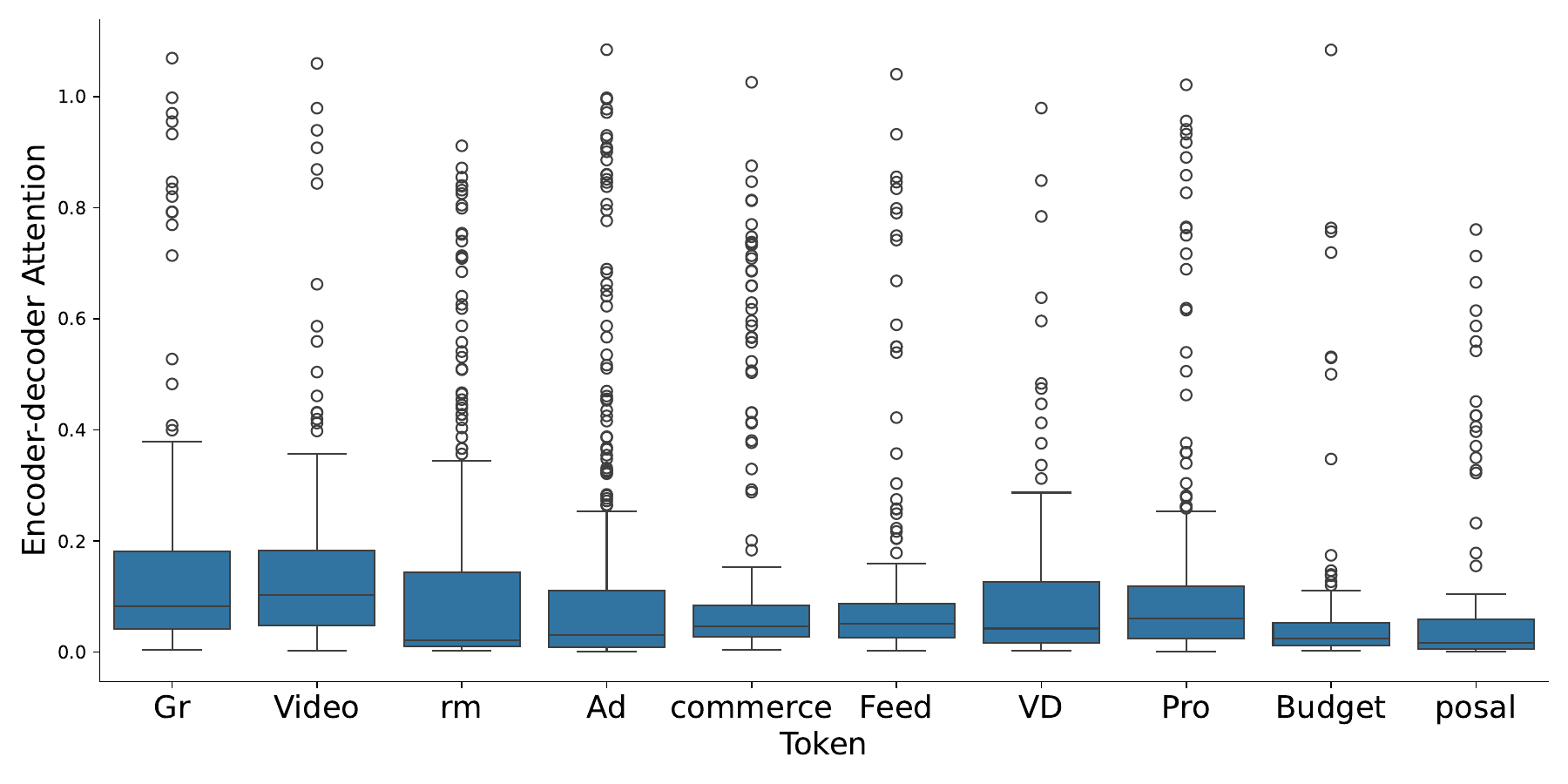}
     \vspace{-16pt}
     \caption{Top 10 tokens with highest variance in encoder-decoder attention for code summarization.}
     \label{fig:en-var-hi}
 \end{subfigure}
 \hfill
 \begin{subfigure}[h]{0.49\textwidth}
     \centering
     \includegraphics[width=\textwidth]{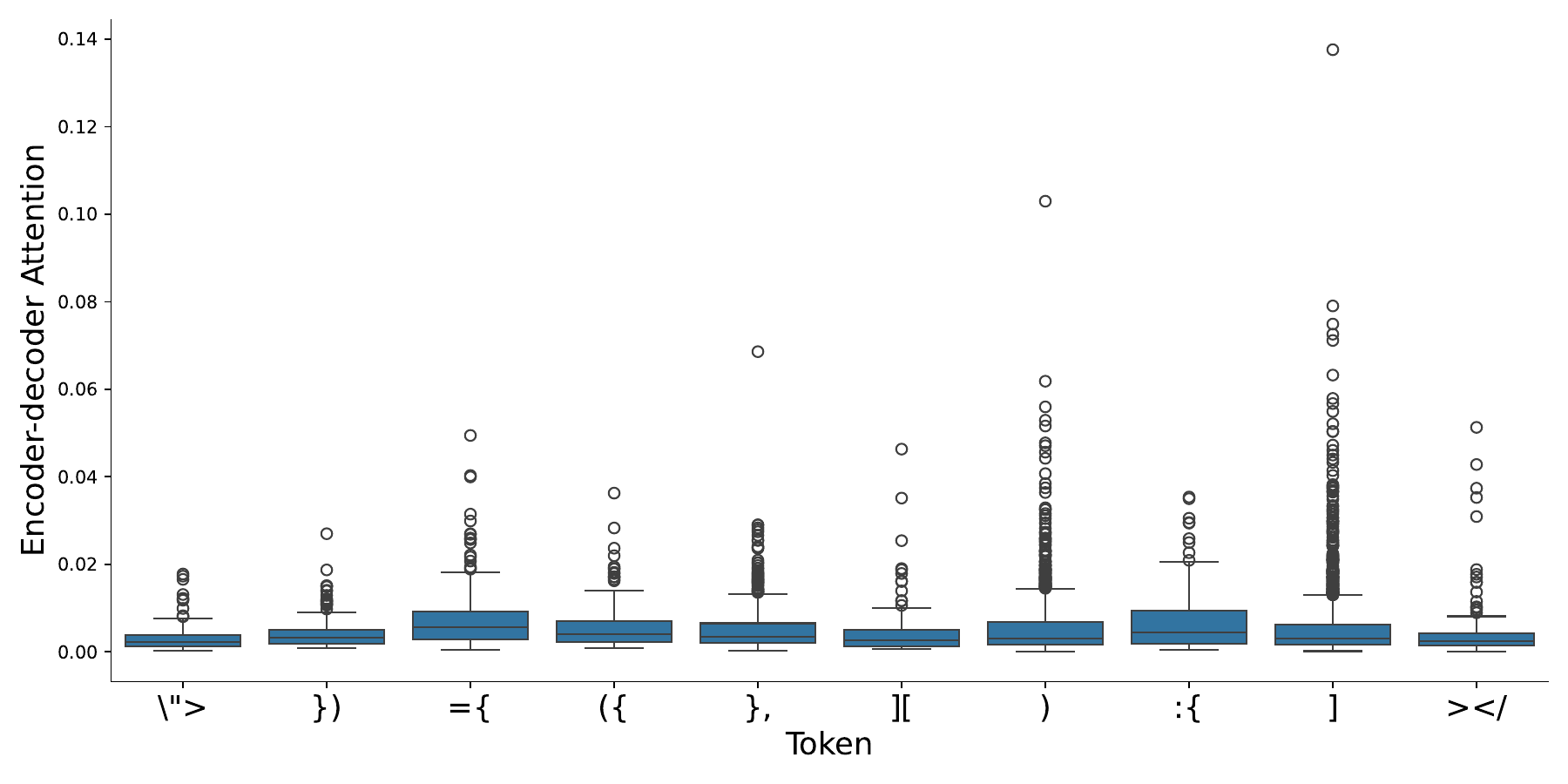}
     \vspace{-16pt}
     \caption{Bottom 10 tokens with lowest variance in encoder-decoder attention for code summarization.}
     \label{fig:en-var-lo}
 \end{subfigure}
 \vspace{-8pt}
    \caption{The variance of the top and bottom 10 tokens of CodeBERT for code search and summarization.}
    \label{fig:var}
\end{figure*}

{\bf (RQ-1) What important tokens do CLS attentions emphasize on?} 
For CodeBERT on code search and summarization, Fig.~\ref{fig:var} show the top 10 and bottom 10 tokens with the highest and lowest variances in attention scores provided by CodeBERT. As seen, the tokens with the highest variance typically encompass richer semantic meanings, such as `accumulate', `pure', and `commerce'. Conversely, tokens with the lowest variance tend to resemble simple symbols, including numbers, separators, and brackets. Upon examining tokens with high variance, we refer back to the original source code to analyze their positions. It is observed that {\em the tokens receive high attention scores when they appear in method signatures, function invocations, or as variables in return statements. Conversely, they attain low scores when used in conditions, expressions, and similar contexts}. Given the substantial variance observed, {\em relying solely on the average of all of its attention scores to gauge a token's importance across diverse statement types is not reasonable}. The attention scores of individual tokens are notably influenced by their positions in the code. Thus, categorizing statements and calculating the average attention score of each token in its contexts, i.e., distinct categories of statements, should be employed, named {\bf context-aware, category-local attention average.} This aims to diminish such variance and bolster accuracy.

\begin{table*}
\centering
\caption{{(RQ-2)} \color{black}{Statistics of encoder-decoder attention scores based on 0.16M training dataset. (Max/Min: the maximum/minimum of encoder-decoder attention scores in each category; Global/Global\_variance: the average/variance of the global attention scores of tokens for each category; Category-local/Local\_variance: the averages/variance of category-local attention scores.)}}
\vspace{-4pt}
\small
\tabcolsep 3.5pt
\begin{tabular}{ccccccc}
\hline
Category & Max & Min & {\bf Global} & {\bf Global\_variance} & {\bf Category-local} & {\bf Local\_variance} \\ 
\hline
Annotation            &  7.94 &  0.32 &  2.61 & 13.76 &  1.55 &  0.09 \\
Arithmetic            & 37.44 &  0.07 &  2.69 & 15.41 &  2.30 &  2.52 \\
Variable Declaration  & 65.54 &  0.09 &  2.87 & 15.63 &  2.69 &  7.97 \\
Function Invocation   & 63.97 &  0.01 &  2.86 & 15.94 &  2.80 &  8.54 \\
Return                & 55.23 &  0.10 &  3.08 & 17.61 &  4.76 & 16.02 \\
Switch                & 30.03 &  0.07 &  2.71 & 16.36 &  2.41 &  2.63 \\
Break                 & 28.02 &  0.04 &  2.64 & 16.43 &  2.67 &  1.21 \\
Setter                & 69.06 &  0.03 &  2.85 & 17.25 &  2.33 &  5.10 \\
Synchronized          & 78.09 &  0.04 &  2.84 & 17.08 &  3.11 &  3.03 \\
Try                   & 78.27 &  0.03 &  2.82 & 17.31 &  2.46 &  2.69 \\
Catch                 & 34.99 &  0.07 &  3.01 & 19.80 &  2.44 &  4.18 \\
Method Signature      & 91.69 &  0.14 &  3.29 & 18.21 &  5.91 & 30.92 \\
Finally               & 10.49 &  0.74 &  2.38 &  7.78 &  2.99 &  1.74 \\
Getter                & 68.49 &  0.03 &  2.88 & 16.59 &  2.58 &  6.42 \\
Throw                 & 87.67 &  0.06 &  2.80 & 16.04 &  3.10 &  8.13 \\
Case                  & 23.25 &  0.03 &  2.75 & 16.11 &  1.80 &  1.55 \\
While                 & 67.68 &  0.04 &  2.70 & 15.52 &  2.41 &  3.14 \\
Continue              &  9.85 &  0.27 &  2.49 & 12.64 &  1.73 &  0.37 \\
If Condition          & 57.88 &  0.05 &  2.84 & 15.84 &  2.50 &  5.97 \\
For                   & 60.62 &  0.03 &  2.91 & 17.21 &  2.99 &  6.89 \\
Logging               & 65.63 &  0.04 &  2.77 & 15.53 &  2.89 &  8.42 \\
\hline
\end{tabular}
\label{tab:en-de-var}
\vspace{-10pt}
\end{table*}

{\bf (RQ-2) What tokens encoder-decoder attentions emphasize on?} 
Table~\ref{tab:en-de-var} shows the average of the Encoder-Decoder attention scores of tokens based on statement classes~\cite{dietcode-fse22}. Unlike CLS attention, each token in the input can have multiple Encoder-Decoder (EnDe) attention scores, i.e., for each generated token, the decoder will calculate an attention score for each token in the input. 
Thus, the largest attention score is selected as the attention score. 
The EnDe attention scores are generated in conjunction with the description. In the instances where the description contains intricate function details, these tokens garner high attention scores, facilitating the establishment of bi-modal mappings. For code search, the significance of detailed information within the code (e.g., \code{Throw} statements) is lower compared to the broader functional description (e.g., `Method signature').


{\bf (RQ-3) Do the averages of self-attention scores reflect
the CLS attentions and the Encoder-Decoder attentions?}
\label{subsubsec:diff}
Our answer is `No'.
{\em The accumulated attention scores from the self-attention (as used in $\toolbase$) is for pre-training and cannot reflect and substitute for those from the CLS and Encoder-Decoder attentions. i.e., the self-attention is used for pre-trained tasks and vectored general representations, not directly for downstream tasks}. For elaboration, these attention schemes are for different tasks. The self attention is for pre-training tasks, while CLS attention is for fine-tuning downstream discriminative tasks, and the Encoder-Decoder attention is for downstream sequence-to-sequence generation tasks. In fact, the encoders of CodeBERT and CodeT5 have been trained in multiple pre-trained tasks. 
%
Thus, the averages of self-attention scores cannot replace the CLS and Encoder-Decoder attentions.
The latter attentions are directly applied to downstream tasks.

\section{\tool: Code Simplification}
\label{sec:algorithm}




\subsection{Code Simplification Problem Formulation}

Given a dataset $D=\{d_1,...,d_{m}\}$ with $m$ snippets. Each snippet $d_j$ contains a sequence of $n_j$ tokens.
Thus, each code snippet $d_j$ can be denoted as $d_j = \{t_1,\cdots,t_{n_j}\}$ and the index of each token records its position.
$w_i$ denotes the importance of each token $t_i$ and $x_i$ is a binary indicator showing whether $t_i$ should be removed or not.
With the simplification ratio $SimplifiedRatio$, the total number of tokens to be removed for $d_j$ is $\mathcal{X} = SimplifiedRatio \times  n_j$.
Now, we formulate code simplification as the combinatorial optimization problem as following:
\vspace{-5pt}
\begin{equation}
\vspace{-5pt}
\begin{split}
& \text{minimize} \sum_{i=1}^{n_j} w_ix_i, \ \text{such that} \sum_{i=1}^{n_j} x_i = \mathcal{X}.
\end{split} 
\label{def:kp}
\end{equation}
\vspace{-10pt}

$ x_i \in \{0,1\}$, code simplification aims to minimize the weighted sum of $w_i x_i$ that satisfies the number of tokens to be removed for each $d_j$.

\subsection{{\tool} Algorithm}


\subsubsection{Computation} As mentioned, there are three methods to measure token importance through attention scores: {\em dynamic, global, and category-local methods}. Regarding the dynamic method,
the CLS and Encoder-Decoder attention scores of the same token will be different for different inputs, which are dynamically generated and can reflect the importance of corresponding tokens in context. 
However, it is inefficient and impractical to assign a dynamic attention score to each input token in the test dataset using the models. Calculating the dynamic attention scores requires multiple transformer blocks (for example, CodeBERT and the encoder of CodeT5 have a stack of 12 transformer blocks), making it time-consuming. Moreover, once we have obtained the dynamic attention scores, the downstream tasks are nearly complete. Thus, it does not make sense and is unnecessary to reduce the code and redo the downstream tasks.

Regarding the global attention average of each token in the training dataset, DietCode uses it to replace the dynamic attention score in testing dataset, which is computed in Equation (\ref{equ:global}):
\vspace{-5pt}
\begin{equation}
\vspace{-5pt}
   \mu_t = \frac{\sum_{j=1}^m \sum_{t \in d'_j} s_t}{n_t}
\label{equ:global}
\vspace{-5pt}
\end{equation}
where $t \in d'_j$ means that a token $t$ is in $d'_j$ in a training dataset $D'$ and $s_t$ is the common self-attention score. 
$n_t$ is the number of the occurrence of token $t$ in the training dataset $D'$.

Our empirical study reveals that significant tokens often have high variances in global attention averages. As a result, we propose the category-local attention average for each token, defined as
\begin{equation}
    \mu^c_t = \frac{\sum_{j=1}^m \sum_{t \in p_k, p_k \in d'_j, L(p_k) \in c} s_t}{n^c_t}.
    \label{equ:local}
\end{equation}
In this equation, $p_k$ is a statement of a code snippet $d'_j$. $ L(p_k)$ is the label (category) of the statement $p_k$. $n^c_t$ refers to the number of occurrences of token $t$ in the statements belonging to the category $c \in C$. Finally, $s_t$ can be CLS attention or Encoder-Decoder attention scores. The definitions of those scores are provided in Section~\ref{subsec:clsendeattion}.


\subsubsection{Removal Algorithm} 
\label{sec:algo}

Algorithm~\ref{alg:leancode} displays {\tool} algorithm.
Unlike DietCode's removing the less critical statements and proceeding to remove less important tokens from other statements, it exclusively focuses on token-level removal, without initially discarding entire statements. 
Deleting entire statements would result in the loss of important tokens. 
Our Algorithm (\ref{alg:leancode}) initializes a copy of the original dataset $D$ as the returned simplified dataset (line 1). Next, it iteratively removes the tokens of each snippet in $D^c$ one by one (lines 2--9) based on their attention scores stored in the dictionary $S = \{t, c, \mu^c_t\}$ where $c$ is the category of the statement that token $t$ belongs to, $\mu^c_t$ is the category-local attention average of the token. At line 3, \code{removedTokens} records the pair of the index and the respective token with current lowest score (\{index:token\}) in $d^c_j$ at each turn. 
The number of removed tokens is set (line 4). In line 5, \code{removedTokenNum} is the number of currently selected tokens to be removed. At each iteration, {\tool} repeatedly selects the remaining token (not in \code{removedTokens}) with the lowest score in $d^c_j$ until the number of removed tokens is reached (lines 6--8). Finally, our algorithm returns the simplified dataset $D'$.

\begin{algorithm}[t]
\caption{\tool: Code Simplification Algorithm}
\label{alg:leancode}
\begin{flushleft}
  \textbf{INPUT:} A dataset $D = \{d_1,\dots,d_m\}$, token scores $S$, $SimplifiedRatio$\\
  \textbf{OUTPUT:} A simplified code dataset $D'$\\
  \textbf{PROCEDURE:}
\end{flushleft}
\begin{algorithmic}[1]
  \State $D^c \gets D$
  \For{$j = 1$ \textbf{to} $m$}
    \State $removedTokens \gets \{\}$
    \State $\mathcal{X} \gets SimplifiedRatio \times n_j$
    \State $removedTokenNum \gets 0$
    \While{$removedTokenNum < \mathcal{X}$}
      \State Add \{index:token with lowest $s_t$\} ($\in d^c_j$, $\notin removedTokens$)  into $removedTokens$
      \State $removedTokenNum$ updates
    \EndWhile
    \State $d^c_j = d^c_j / removedTokens[1:\mathcal{X}]$
  \EndFor
  \State \textbf{return} $D^c$
\end{algorithmic}
\end{algorithm}

\section{Empirical Evaluation}
\label{sec:eval}
We evaluate our \tool on different code tasks using three PLLMs and GPT-4o.

\noindent {\bf Downstream Tasks}: We choose code search and summarization as the tasks (also used in \toolbase and \toolbasebase), which are commonly used in evaluating LLMs in text and code analysis~\cite{ahmed2022multilingual,codebert-emnlp20,jiang2021treebert,wang2021syncobert,liu2022deeplearning,liu2019roberta}. The goal of code search is to find relevant code snippets from a codebase given a query and code summarization is to generate a natural language summary for a given code. 





\noindent {\bf Models Under Study}: 
We opted for 3 popular models, CodeBERT~\cite{codebert-emnlp20}, CodeT5~\cite{wang-etal-2021-codet5} and GPT-4o~\cite{islam2024gpt}. Unlike {\toolbase} (calculating attention scores based on all layers), for `CLS' and Encoder-Decoder attention scores, we obtain them from the last encoder and decoder blocks of the respective model.




{\em \bf CodeBERT-based code search and summarization:} 
We added a fully connected layer on top of the CodeBERT model for binary classification to perform code search. As in DietCode, we added a Transformer decoder for code summarization. 
    
{\em \bf CodeT5-based code search and summarization:} For code search, its encoder is separately extracted and joint with a fully connected layer for the classification task. We use CodeT5 directly for code summarization.





{\color{custom-blue}{{\em \bf GPT4-based code search and summarization:} Since GPT-4o is accessible only through programming APIs, we cannot directly access its model. 
We are limited to using prompts to obtain classification and summrization results and corresponding analyses from GPT-4o in a predefined format.
\noindent {\bf Baselines}: We chose {\bf {\toolbase}}~\cite{dietcode-fse22} and {\bf {\toolbasebase}}~\cite{wang2024natural}, the SOTA code simplification methods. DietCode~\cite{dietcode-fse22} is based on token weights learned by models and SlimCode~\cite{wang2024natural} is based on the nature of tokens.



\noindent {\bf Datasets}: We used code search and summarization datasets from CodeBERT~\cite{codebert-emnlp20} (Details in Table~\ref{tab:statistics_code} in Appendix). These are the extensions of CodeSearchNet~\cite{husain2019codesearchnet}, which is a collection of datasets and benchmarks for code retrieval using texts. It consists of +2 millions pairs of (comment, code) that were extracted from Github, covering six languages (Python, PHP, Go, Java, JavaScript, and Ruby). Since DietCode~\cite{dietcode-fse22} reported similar trends for different languages, we
conducted experiments only on Java.

\noindent {\bf Metrics}: We use the simplification ratio to measure the degree of simplification of a code snippet. Given a code snippet $Code$ and its simplified one $Scode$, $SimplifiedRatio = \frac{|Code|-|Scode|}{|Code|} \times 100$.
$|Code|$ and $|Scode|$ are the numbers of tokens in $Code$ and $Scode$. 
The efficiency of simplified code is measured by the \emph{\textbf{time cost}} it takes for model inference.
We use {{\em \textbf{BLEU-4}}} score and {\em \textbf{MRR}} (Mean Reciprocal Rank) for code summarization and search, respectively. 
For code search by GPT-4o, we use Precision instead of MRR due to the latter's high computational requirements. \emph{Precision} is highly correlated with MRR in measuring model effectiveness. Similar to $\toolbasebase$, we randomly replace the code description in 400 sample pairs of texts-code and check if the replacement content matches the code part. 
The dataset consists of an equal number of matching and non-matching samples.

%




\noindent {\bf Implementation}: 
DietCode is realized using the code in~\cite{dietcode-fse22}.
We set up CodeBERT and CodeT5 with default hyper-parameters. 
For optimization, they used Adam optimizer with learning rates of $1 \times 10^{-5}$ and $5 \times 10^{-5}$ for downstream tasks. We used a server with 2 CPUs of Intel(R) Xeon(R) Golden 2.40GHz and 2 Nvidia A100s.
\vspace{-6pt}


\begin{table*}[t]
    \centering
    \footnotesize
    \tabcolsep 1.2pt
    \caption{Results of \underline{Code Search} for $\toolbase$ , SlimCode and $\tool$. (10\%-50\%: removing 10\%-50\% tokens, R-M: Reduced MRR, Time: Inference time, R-T: Reduced Inference time)}
    \vspace{-8pt}
    \begin{tabular}{l|cc|cc|cc|cc|cc|cc|cc|cc}
    \hline
    \multirow{3}{*}{\textbf{Ratio}} & \multicolumn{4}{c|}{\textbf{$\toolbase$}} & \multicolumn{4}{c|}{\textbf{SlimCode}} & \multicolumn{4}{c|}{\textbf{$\tool$}} & \multicolumn{4}{c}{\textbf{Inference}} \\
    \cline{2-17}
    & \multicolumn{2}{c|}{\textbf{CodeBERT}} & \multicolumn{2}{c|}{\textbf{CodeT5}} & \multicolumn{2}{c|}{\textbf{CodeBERT}} & \multicolumn{2}{c|}{\textbf{CodeT5}} & \multicolumn{2}{c|}{\textbf{CodeBERT}} & \multicolumn{2}{c|}{\textbf{CodeT5}} & \multicolumn{2}{c|}{\textbf{CodeBERT}} & \multicolumn{2}{c}{\textbf{CodeT5}} \\
    & MRR & R-M & MRR & R-M& MRR & R-M & MRR & R-M& MRR & R-M & MRR & R-M & Time & R-T & Time & R-T    \\
    \hline
    Base & 0.726 &  --- & 0.747 & --- & 0.726 & --- & 0.747 & --- & 0.726 & --- & 0.747 & --- & 41m & --- & 40m & --- \\
    10\% & 0.663&  8.67\%\color{green}{↓}& 0.699& 6.42\%\color{green}{↓}& 0.731& 0.68\%\color{red}{$\uparrow$}& 0.738& 1.2\%\color{green}{↓}& 0.728& 0.27\%\color{red}{$\uparrow$}& 0.743& 0.53\%\color{green}{↓}& 38m& 7.31\%\color{green}{↓}& 36m& 10\%\color{green}{↓}
\\
    20\% & 0.598&  17.63\%\color{green}{↓}& 0.669& 10.44\%\color{green}{↓}& 0.726& 0.00\%\color{green}{↓}& 0.733& 1.87\%\color{green}{↓}& 0.719& 0.96\%\color{green}{↓}& 0.736& 1.47\%\color{green}{↓}& 35m& 14.63\%\color{green}{↓}& 33m& 17.5\%\color{green}{↓}
\\
    30\% & 0.529&  27.13\%\color{green}{↓}& 0.624& 16.46\%\color{green}{↓}& 0.70& 3.58\%\color{green}{↓}& 0.723& 3.21\%\color{green}{↓}& 0.716& 1.37\%\color{green}{↓}& 0.724& 3.07\%\color{green}{↓}& 32m& 21.95\%\color{green}{↓}& 31m& 22.5\%\color{green}{↓}
\\
    40\% & 0.502&  30.85\%\color{green}{↓}& 0.602& 19.41\%\color{green}{↓}& 0.632& 12.94\%\color{green}{↓}& 0.679& 9.1\%\color{green}{↓}& 0.697& 3.99\%\color{green}{↓}& 0.714& 4.41\%\color{green}{↓}& 29m& 29.27\%\color{green}{↓}& 28m& 30\%\color{green}{↓}
\\
    50\% & 0.429&  40.90\%\color{green}{↓}& 0.561& 24.89\%\color{green}{↓}& 0.594& 18.18\%\color{green}{↓}& 0.641& 14.19\%\color{green}{↓}& 0.688& 5.23\%\color{green}{↓}& 0.706& 5.48\%\color{green}{↓}& 26m& 36.59\%\color{green}{↓}& 25m& 37.5\%\color{green}{↓}\\
    \hline 
    \end{tabular}
  \label{tab:CodeSearch}
\end{table*}

\begin{table*}[t]
    \centering
    \footnotesize
    \tabcolsep 0.4pt
    \caption{Results of \underline{Code Summarization} for $\toolbase$ , SlimCode and $\tool$. (10\%-50\%: removing 10\%-50\% tokens, R-B: Reduced BLEU, Time: Inference time, R-T: Reduced Inference time)} 
    \vspace{-8pt}
    \begin{tabular}{l|cc|cc|cc|cc|cc|cc|cc|cc}
    \hline
    \multirow{3}{*}{\textbf{Ratio}} & \multicolumn{4}{c|}{\textbf{$\toolbase$}} & \multicolumn{4}{c|}{\textbf{SlimCode}} & \multicolumn{4}{c|}{\textbf{$\tool$}} & \multicolumn{4}{c}{\textbf{Inference}} \\
    \cline{2-17}
    & \multicolumn{2}{c|}{\textbf{CodeBERT}} & \multicolumn{2}{c|}{\textbf{CodeT5}} & \multicolumn{2}{c|}{\textbf{CodeBERT}} & \multicolumn{2}{c|}{\textbf{CodeT5}} & \multicolumn{2}{c|}{\textbf{CodeBERT}} & \multicolumn{2}{c|}{\textbf{CodeT5}} & \multicolumn{2}{c|}{\textbf{CodeBERT}} & \multicolumn{2}{c}{\textbf{CodeT5}} \\
    & BLUE & R-B & BLUE & R-B& BLUE & R-B & BLUE & R-B& BLUE & R-B & BLUE & R-B & Time & R-T & Time & R-T    \\
    \hline
    Base & 18.25&  --- & 20.55& --- & 18.25& --- & 20.55& --- & 18.25& --- & 20.55& --- & 29m& --- & 22m& --- \\
    10\% & 16.44&  9.91\%\color{green}{↓}& 17.27& 15.96\%\color{green}{↓}& 17.86& 2.13\%\color{green}{↓}& 20.01& 2.62\%\color{green}{↓}& 18.08& 0.93\%\color{green}{↓}& 20.32& 1.11\%\color{green}{↓}& 27m& 6.89\%\color{green}{↓}& 20m& 9.09\%\color{green}{↓}\\
    20\% & 15.68&  14.08\%\color{green}{↓}& 16.48& 19.80\%\color{green}{↓}& 17.35& 4.93\%\color{green}{↓}& 18.68& 4.23\%\color{green}{↓}& 17.73& 2.84\%\color{green}{↓}& 20.18& 1.80\%\color{green}{↓}& 26m& 10.34\%\color{green}{↓}& 18m& 18.18\%\color{green}{↓}\\
    30\% & 15.05&  17.53\%\color{green}{↓}& 15.74& 23.40\%\color{green}{↓}& 16.8& 7.94\%\color{green}{↓}& 18.74& 8.80\%\color{green}{↓}& 17.23& 5.58\%\color{green}{↓}& 19.82& 3.55\%\color{green}{↓}& 24m& 17.24\%\color{green}{↓}& 16m& 27.27\%\color{green}{↓}\\
    40\% & 14.66&  19.67\%\color{green}{↓}& 15.11& 26.47\%\color{green}{↓}& 15.95& 12.60\%\color{green}{↓}& 16.35& 20.43\%\color{green}{↓}& 16.71& 8.43\%\color{green}{↓}& 19.27& 6.22\%\color{green}{↓}& 23m& 20.68\%\color{green}{↓}& 15m& 31.81\%\color{green}{↓}\\
    50\% & 14.23&  22.02\%\color{green}{↓}& 14.27& 30.55\%\color{green}{↓}& 15.28& 16.27\%\color{green}{↓}& 14.53& 29.29\%\color{green}{↓}& 16.24& 11.01\%\color{green}{↓}& 18.46& 10.17\%\color{green}{↓}& 23m& 24.13\%\color{green}{↓}& 13m& 40.90\%\color{green}{↓}\\
    \hline 
    \end{tabular}
  \label{tab:CodeSummarization}
\end{table*}

\subsection{\bf Effectiveness of {\tool}}\label{RQ4}

Tables~\ref{tab:CodeSearch} and~\ref{tab:CodeSummarization} present the results. We first use the training dataset (complete code snippets) to train the relevant models, and then use the testing dataset (complete code snippets) to perform inference on a model to obtain the `Base' results. Next, we apply code simplification to the samples in the testing dataset using DietCode, SlimCode, and {\tool} according to the simplifiedRatio. Finally, we input the simplified code into the trained models to perform inference, obtaining results for the cases of the removal percentages of 10\%-50\%.
As seen in columns `R-M' and `R-B', the performance of both methods declines in downstream tasks as the $SimplifiedRatio$ gradually increases. However, even when 50\% of the code is removed, the decline at most in the effect of $\tool$ for code search and code summarization are only {\em 5.48\%} and {\em 11.01\%}. Meanwhile, \tool enhances \toolbase (\toolbasebase) up to 60.37\% (15.82\%) and 14.12\%(6.28\%) in terms of MRR and BLEU scores with CodeBERT. Similarly, it improves by 25.84\% (10.14\%) and 29.36\% (27.04\%) in terms of MRR and BLEU scores with CodeT5.

We observe two interesting points: 1) Intuitively, one would expect the performance of the removed code for downstream tasks to be worse than the original code. However, \tool and \toolbasebase can achieve better results on code search. 2) Compared with \tool, \toolbasebase experiences a sharp decline in performance when the $simplifiedRatio$ exceeds 30\%.

For the $1^{st}$observation, if the code contains more valuable information, the downstream tasks will have better results. Not all code snippets can be fully inserted into models, as the maximum lengths for original codes are limited to 512 tokens.
Thus, even if some tokens are removed, tokens towards the end of the code snippets may still be included in the input.
If the newly input tokens provide more information than the removed tokens, then the above observation may occur. Specifically, {\tool} often removes low-quality tokens with lower attention scores (e.g., symbol-like tokens) while allowing high-quality tokens to enter the model input. 

For the $2^{nd}$ observation, the main reason lies in the deletion of increasingly important tokens as the process progresses. At this stage, a model's ability to precisely discern token importance becomes critical for maintaining the task performance. \toolbasebase, with only 8 levels of token importance, struggles to differentiate between crucial tokens when $simplifiedRatio$ surpasses 30\%.

Figure~\ref{fig:demonstration} presents an example of our \tool. Subfigure (a) shows the original code, which computes a Bessel function using a recurrence formula. Subfigures (b) and (c) display the results of \tool after 30\% token removal for code search and summarization, respectively. Both approaches preserved critical elements, such as function signatures and return statements. For code search, structural details like loops, arithmetic operations, and variable names were retained to support precise keyword matching. For summarization, non-essential details such as loop bodies, variable declarations, and calculations were removed, while return statements were emphasized. This demonstrates how return statements play a key role in capturing code intent, enabling concise and clear summaries.

\begin{figure*}[t]
  \centering
  \includegraphics[width=\textwidth,height=6cm]{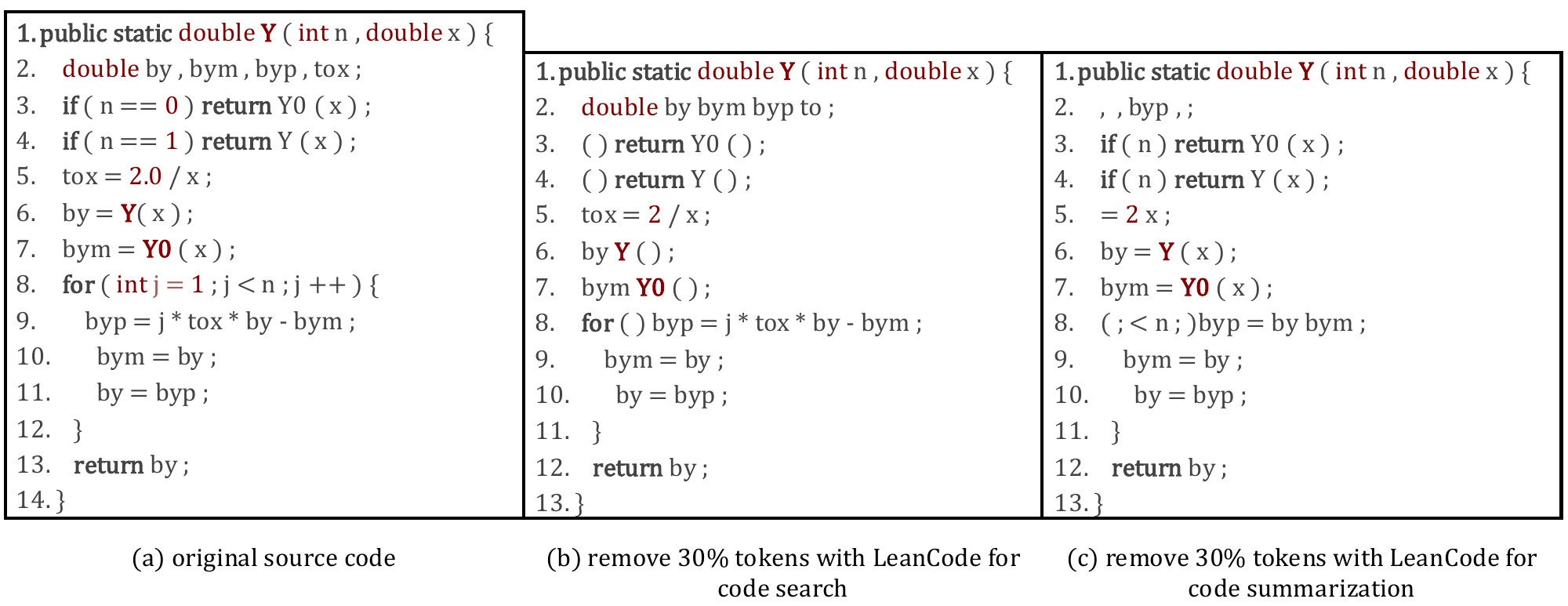}
  \vspace{-18pt}
  \caption{The example of \tool for code simplification.}
  \label{fig:demonstration}
\end{figure*}

\subsection{\bf Efficiency of \tool}

Tables~\ref{tab:CodeSearch} and~\ref{tab:CodeSummarization} show the results on the positive correlations between the simplified ratio and inference time.  
For instance, 
for code summarization using CodeT5, {\em the inference time is reduced by 40.9\% when the simplification ratio is set to 50\%}. 

For both tasks, as the simplified ratio increases, {\em the inference time follows a near-linear descent} on both of models. Specifically, for code search, the ratio of $SimplifiedRatio$ and reduce time is about 0.7, meanwhile, for code summarization, the ratio is about 0.5 on CodeBERT and 0.75 on CodeT5, respectively.
The ratio on CodeBERT for code summarization is relative lower than others since directly concatenating CodeBERT with a Transformer decoder would result in the model lacking some optimization for inference.

In addition to inference time, each method requires training and pruning time. DietCode and LeanCode also need extra computation for per-token attention scores. While all methods share identical training times due to using the same training dataset  and model, calculating token attention scores adds about 5\% to the total training time. For example, in code summarization, standard training across 8 epochs takes 37.5 minutes per epoch, totaling 300 minutes. When attention scores are collected only during the final epoch, the first seven epochs remain unchanged, but the last epoch increases to 53 minutes. This results in a total time of 315.5 minutes,i.e., representing a 5\% increase in training time.

We conducted additional pruning experiments for LeanCode and obtained pruning time results for SlimCode and DietCode from SlimCode's paper. The results are shown in Table~\ref{tab:pruning_times}. From these results, we observed: 1) DietCode's pruning time is significantly higher than SlimCode and LeanCode due to its two-step process—selecting high-quality statements first, then removing tokens from the remaining ones. This approach leads to complex knapsack optimization problems, resulting in a lower reduction ratio and increased pruning time; 2) LeanCode's pruning time is approximately 2–4 times longer than SlimCode's, primarily due to LeanCode's more complex pruning process, which includes additional steps like tokenizing and statement class matching; 3) The pruning times for LeanCode and SlimCode remain within a comparable range.

\begin{table}[ht]
  \centering
  \small
  \caption{Pruning times for DietCode(\textbf{DC}), SlimCode(\textbf{SC}), and LeanCode(\textbf{LC}) on code search and code summarization training datasets(10\%-50\% removal for each code snippet)}
  \label{tab:pruning_times}
  \vspace{-6pt}
  \setlength{\tabcolsep}{3pt}
  \renewcommand{\arraystretch}{0.8}
  \begin{tabular}{c|ccc|ccc}
    \hline
    \multirow{2}{*}{\textbf{Ratio}} 
      & \multicolumn{3}{c|}{\textbf{Code Search}} 
      & \multicolumn{3}{c}{\textbf{Code Summarization}} \\
      & \textbf{DC} & \textbf{SC} & \textbf{LC}
      & \textbf{DC} & \textbf{SC} & \textbf{LC} \\
    \hline
    10\% & 9h24m   & 17m    & 46m33s & 1h40m & 45s  & 3m32s \\
    20\% & 8h28m   & 17m    & 46m39s & 1h30m & 53s  & 3m37s \\
    30\% & 7h37m   & 20m    & 47m15s & 1h19m & 59s  & 3m41s \\
    40\% & 6h45m   & 21m    & 47m35s & 1h11m & 66s  & 3m45s \\
    50\% & 5h59m   & 21m    & 47m43s & 1h02m & 69s  & 3m50s \\
    \hline
  \end{tabular}
\end{table}

\begin{table*}[t]
    \centering
    \footnotesize
    \setlength{\tabcolsep}{7.7pt} 
    \caption{Results of removal methods on GPT-4 for \underline{Code Search} (IT: Input Tokens, R-IT: Reduced Input Tokens (\%), OT:Output Tokens, R-OT:Reduced Output Tokens (\%); TT:Total Tokens, R-P:Reduced Precision (\%)} 
    \vspace{-2mm} 
    \begin{tabular}{l|cc|cc|cc|cc}
    \hline 
    \textbf{Removal method} & \textbf{IT} & \textbf{R-IT} & \textbf{OT} & \textbf{R-OT} & \textbf{TT} & \textbf{R-TT} & \textbf{Precision} & \textbf{R-P} \\
    \hline 
    Base & 102385 & --- & 24535 & --- & 127920 & --- & 0.82 & --- \\
    \hline 
    DietCode (10\%) & 98105& 4.18\%\color{green}{↓}& 24619 & 0.34\%\color{red}{↑}& 122724& 4.06\%\color{green}{↓}& 0.776 & 5.37\%\color{green}{↓}\\
    DietCode (20\%) & 93268 & 8.23\%\color{green}{↓} & 24508 & 0.11\%\color{green}{↓} & 117776 & 7.92\%\color{green}{↓} & 0.813 & 0.85\%\color{green}{↓}\\
    DietCode (30\%) & 86672 & 14.61\%\color{green}{↓} & 24211 & 1.32\%\color{green}{↓} & 110883 & 13.32\%\color{green}{↓} & 0.775 & 5.49\%\color{green}{↓}\\
    DietCode (40\%) & 80196 & 20.98\%\color{green}{↓} & 24536 & 0.00\%\color{red}{↑}& 104732 & 18.14\%\color{green}{↓} & 0.78 & 4.88\%\color{green}{↓}\\
    DietCode (50\%) & 73759 & 27.76\%\color{green}{↓} & 24169 & 1.49\%\color{green}{↓} & 97928 & 23.44\%\color{green}{↓} & 0.776 & 5.37\%\color{green}{↓}\\
    \hline 
    SlimCode (10\%) & 96845& 5.41\%\color{green}{↓}& 25343 & 3.29\%\color{red}{↑} & 122188& 4.48\%\color{green}{↓}& 0.79 & 3.66\%\color{green}{↓}\\
    SlimCode (20\%) & 91736& 10.40\%\color{green}{↓}& 25217 & 2.99\%\color{red}{↑}& 116953& 8.57\%\color{green}{↓}& 0.808 & 1.46\%\color{green}{↓}\\
    SlimCode (30\%) & 85778& 16.22\%\color{green}{↓}& 24779 & 0.99\%\color{red}{↑}& 110557& 13.57\%\color{green}{↓}& 0.789 & 3.78\%\color{green}{↓}\\
    SlimCode (40\%) & 79686& 22.17\%\color{green}{↓}& 24734 & 0.81\%\color{red}{↑}& 104420& 18.37\%\color{green}{↓}& 0.796 & 2.93\%\color{green}{↓}\\
    SlimCode (50\%) & 73215& 28.49\%\color{green}{↓}& 24548 & 0.05\%\color{red}{↑}& 97763& 23.57\%\color{green}{↓}& 0.763 & 6.95\%\color{green}{↓}\\
    \hline 
    LeanCode (10\%) & 96978 & 5.28\%\color{green}{↓}& 25843 & 5.33\%\color{red}{↑}& 122821 & 3.99\%\color{green}{↓} & 0.795 & 3.05\%\color{green}{↓}\\
    LeanCode (20\%) & 91173 & 10.95\%\color{green}{↓}& 25043 & 2.07\%\color{red}{↑} & 116216 & 9.15\%\color{green}{↓} & 0.798 & 2.68\%\color{green}{↓}\\
    LeanCode (30\%) & 85202 & 16.78\%\color{green}{↓}& 25454 & 3.74\%\color{red}{↑} & 110656 & 13.50\%\color{green}{↓}& 0.828 & 0.49\%\color{red}{↑}\\
    LeanCode (40\%) & 79102 & 22.74\%\color{green}{↓}& 25162 & 2.55\%\color{red}{↑} & 104264 & 18.49\%\color{green}{↓}& 0.793 & 3.54\%\color{green}{↓}\\
    LeanCode (50\%) & 72887 & 28.81\%\color{green}{↓}& 24741 & 0.84\%\color{red}{↑}& 97628 & 23.68\%\color{green}{↓}& 0.81 & 1.22\%\color{green}{↓}\\
    \hline 
    \end{tabular}
    \label{tab:CodeSearchGPT4}
\end{table*}

\begin{table*}[t]
    \centering
    \footnotesize
    \setlength{\tabcolsep}{9pt}
    \caption{ Results of removal methods on GPT-4 for \underline{Code Summarization} (IT: Input Tokens, R-IT: Reduced Input Tokens (\%), OT: Output Tokens, R-OT: Reduced Output Tokens (\%); TT: Total Tokens, R-B: Reduced BLEU (\%))}
    \vspace{-2mm}
    \begin{tabular}{l|cc|cc|cc|cc}
    \hline 
    \textbf{Removal method} & \textbf{IT} & \textbf{R-IT} & \textbf{OT} & \textbf{R-OT} & \textbf{TT} & \textbf{R-TT} & \textbf{BLEU} & \textbf{R-B} \\
    \hline 
    Base & 78246 & --- & 7668 & --- & 85914 & --- & 10.59 & --- \\
    \hline 
    DietCode (10\%) & 75217& 3.87\%\color{green}{↓}& 7340 & 4.28\%\color{green}{↓} & 82557& 3.90\%\color{green}{↓}& 10.80 & 1.98\%\color{red}{↑} \\
    DietCode (20\%) & 72245 & 7.67\%\color{green}{↓} & 7583 & 1.11\%\color{green}{↓} & 79828 & 7.08\%\color{green}{↓} & 10.21 & 3.59\%\color{green}{↓} \\
    DietCode (30\%) & 66844 & 14.58\%\color{green}{↓} & 7846 & 232\%\color{red}{↑}& 74690 & 13.06\%\color{green}{↓} & 10.12 & 4.44\%\color{green}{↓} \\
    DietCode (40\%) & 61534 & 21.36\%\color{green}{↓} & 8015 & 4.52\%\color{red}{↑} & 69549 & 19.06\%\color{green}{↓} & 9.95 & 6.04\%\color{green}{↓} \\
    DietCode (50\%) & 56162 & 28.21\%\color{green}{↓} & 7748 & 1.04\%\color{red}{↑} & 63910 & 25.59\%\color{green}{↓} & 9.69 & 8.50\%\color{green}{↓} \\
    \hline 
    SlimCode (10\%) & 74912& 4.26\%\color{green}{↓}& 7426 & 3.16\%\color{green}{↓} & 82338& 4.16\%\color{green}{↓}& 10.71 & 1.14\%\color{red}{↑} \\
    SlimCode (20\%) & 70734& 9.96\%\color{green}{↓}& 7715 & 0.61\%\color{red}{↑}& 78449& 8.69\%\color{green}{↓}& 10.89 & 2.83\%\color{red}{↑} \\
    SlimCode (30\%) & 65648& 16.10\%\color{green}{↓}& 7633 & 0.46\%\color{green}{↓} & 73281& 14.70\%\color{green}{↓}& 10.71 & 1.14\%\color{red}{↑} \\
    SlimCode (40\%) & 60585& 22.57\%\color{green}{↓}& 7773 & 1.37\%\color{red}{↑} & 68358& 20.43\%\color{green}{↓}& 10.62 & 0.28\%\color{red}{↑} \\
    SlimCode (50\%) & 55335& 29.28\%\color{green}{↓}& 7496 & 2.24\%\color{green}{↓} & 62831& 26.86\%\color{green}{↓}& 10.60 & 0.09\%\color{red}{↑} \\
    \hline 
    LeanCode (10\%) & 74938 & 4.23\%\color{green}{↓} & 7487 & 2.36\%\color{green}{↓}& 82425 & 4.06\%\color{green}{↓} & 11.11 & 4.91\%\color{red}{↑} \\
    LeanCode (20\%) & 70207 & 10.28\%\color{green}{↓} & 8278 & 7.95\%\color{red}{↑} & 78485 & 8.66\%\color{green}{↓} & 10.69 & 1.09\%\color{red}{↑} \\
    LeanCode (30\%) & 65322 & 16.51\%\color{green}{↓} & 7607 & 0.80\%\color{green}{↓} & 72929 & 15.11\%\color{green}{↓} & 10.77 & 1.70\%\color{red}{↑} \\
    LeanCode (40\%) & 60296 & 22.93\%\color{green}{↓} & 8163 & 6.45\%\color{red}{↑} & 68459 & 20.32\%\color{green}{↓} & 10.90 & 2.93\%\color{red}{↑} \\
    LeanCode (50\%) & 55261 & 29.37\%\color{green}{↓} & 8442 & 10.08\%\color{red}{↑} & 63703 & 25.86\%\color{green}{↓} & 10.70 & 1.04\%\color{red}{↑} \\
    \hline 
    \end{tabular}
    \label{tab:CodeSummarizationGPT4}
\end{table*}
\subsection{Model Transferability}
We assess whether code simplified by \tool using one model can be applied to another while maintaining previous results and conclusions for a given downstream task. We replicate the analysis procedures in the Section~\ref{RQ4} with GPT-4o. 
As GPT-4o only provides access via APIs, we used a program to interact with GPT-4o. We used the following prompts: 1) for code search, {\em ``Please check whether the incomplete code snippet is semantically consistent with the given $text$. Please analyze step-by-step first, and then answer in the following format."}, and 2) for code summarization, {\em ``Write a short sentence to describe the function of the incomplete code snippet. Answer in the following format."} Finally, we analyzed the results of GPT-4o to calculate the number of total tokens, precision, and BLEU-4.


Tables~\ref{tab:CodeSearchGPT4} and~\ref{tab:CodeSummarizationGPT4} present the results of the code simplification methods. We use the total number of tokens (the input and output tokens), instead of the prediction time for two reasons: 1) the fee charged for the usage of computing resource of GPT-4o is based on the total number of tokens~\cite{openai-pricing}; 2) GPT-4 only offers API interfaces, making it difficult to accurately measure time due to multiple influencing factors.
It's worth noting that the input tokens include not only the code snippet but also the prompts.
From the results, we made the following empirical observations:


{\em a) Saving Computational Resources.}
As the simplified ratio increases, the total number of tokens decrease for all methods on both tasks in GPT-4o. This finding substantiates that code simplification leads to a reduction in resources required for Prompt-based LLM. This can be due the fact that GPT-4o requires a significant amount of analytical content by following a sequence of statements through reasoning, as opposed to solely producing binary responses. It shows that in the absence of generating analytical texts, the classification performance of GPT-4o is notably deficient~\cite{DBLP:journals/corr/abs-2303-08774}. 
However, the total number of tokens decreases since the reduction in input tokens outweighs the increase in output tokens.


{\em b) Performance on Code Summarization.}
Firstly, {\tool} still slightly outperforms the baselines. 
We observe that for all methods, the BLEU-4 scores in Table~\ref{tab:CodeSummarizationGPT4}  are approximately half of those in Table~\ref{tab:CodeSummarization}. This indicates that the descriptions from GPT-4o deviate more largely from the ground truth compared to those from CodeT5. GPT-4o lacks the capability to produce descriptions closely resembling the ground truth, as it did not undergo fine-tuning. 
The overlap of words between the generated description and the ground truth is considerably lower when compared to CodeT5. 


{\em c) Performance on Code Search.} From Table~\ref{tab:CodeSearchGPT4}, the base GPT-4o yielded good results, even without model fine-tuning. Interestingly, as the simplified ratio increases, Precision scores do not show a consistent decline. Moreover, all methods produced closely clustered results, with \tool maintaining a slight edge.
A plausible explanation is that GPT-4o was trained on a similar dataset. Consequently, GPT-4o can identify crucial tokens for generating judgments, resulting in better inference. 


\section{Related Work}
\label{sec:related}

{\bf Pre-trained Models.} Pre-trained models 
have significantly advanced code models~\cite{fengetal2020codebert, guo2020graphcodebert,karampatsis2020scelmo,guo2022unixcoder}. They excel in various tasks such as code generation \cite{wang-etal-2021-codet5,clement2020pymt5,wang2023codet5+}, defect detection \cite{wang2023deepvd}, 
code summarization \cite{ahmed2022multilingual,jiang2021treebert}, and code search \cite{wang2021syncobert,liu2022deeplearning}.
%
Researchers have investigated pre-trained models for code understanding \cite{ahmad2021unified,mastropaolo2021studying}. For instance, 
Karmakar and Robbes \cite{karmakar2021pre} conducted four probing tasks on pre-trained models to assess their ability to learn various aspects of source code.
Wan {\em et al.}~\cite{wan2022they} reported that Transformer attention mechanisms can capture high-level structures within source code. Moreover, Autofocus~\cite{bui2019autofocus} is a method to determine the most relevant code by measuring statement relevance using attention weights from a GGNN \cite{DBLP:conf/iclr/AllamanisBK18}.

{\bf Program Simplification}. SIV\-AND \cite{rabin2021understanding} and P2IM \cite{suneja2021probing} typically build upon the delta debugging prototype \cite{zeller2002simplifying}, which involves treating a code snippet and an auxiliary model (e.g., code2vec) as inputs. The model segments the code snippet into fragments and use each as inputs. If a~fragment achieves a satisfactory score, it is further divided. This process continues until the subset's performance fails to meet the desired score, resulting in the smallest snippet that satisfies the~goal. 

\section{Conclusion}

In this study, we introduce {\tool}, a novel code simplification approach that harnesses code contexts to utilize attention scores of pre-trained models for representing the importance levels of each token of input. We advocate for the selective removal of tokens based on the average {\bf context-aware} attention scores.
rather than relying on average scores across all inputs. 
We evaluated {\tool} in code search and code summarization tasks, 
experimental results show its superiority over the SOTA DietCode and SlimCode, achieving improvements of up to 60\% and 16\% for code search and 29\% and 27\% for code summarization. 

\section{Limitations}
\label{sec:threats}
{\tool} still has the following limitations"

{\bf Programming Language}: One limitation of our current model is its exclusive application to Java, which restricts its use and effectiveness in other programming languages. Although related literature reports similar effects in other languages, this limitation highlights the need for future work to expand the model's capabilities across multiple programming languages.

{\bf External Validity}: Our experiments were on three models (CodeBERT, CodeT5, GPT-4o), and two tasks of code search (text-to-code) and code summarization (code-to-text). Despite the consistent findings, for generalizability, future experiments are needed on a wider variety of models with other paradigms and in other code-related tasks. Our dataset might not be representative. However, for a fair comparison, we used the same datasets and two tasks as in \toolbase and \toolbasebase.

{\bf Internal Validity}: When measuring the time, we acknowledge that there might be other external factors involving hardware (e.g., GPUs), operating system delays, etc. However, inference time for the models, CodeBERT and CodeT5 is stable in the same controlled environment. 



\section*{Acknowledgments}
The third author, Tien N. Nguyen, was supported in part by the US National Science Foundation (NSF) grant CNS-2120386 and the National Security Agency (NSA) grant NCAE-C-002-2021.

\bibliography{acl_latex}

\section*{Appendix}

\renewcommand{\thesubsection}{\Alph{subsection}}

\subsection{DietCode discarded encoder-decoder attention scores}

\begin{figure*}
\centering
\begin{subfigure}[b]{0.48\textwidth}
   \centering
   \includegraphics[width=\textwidth]{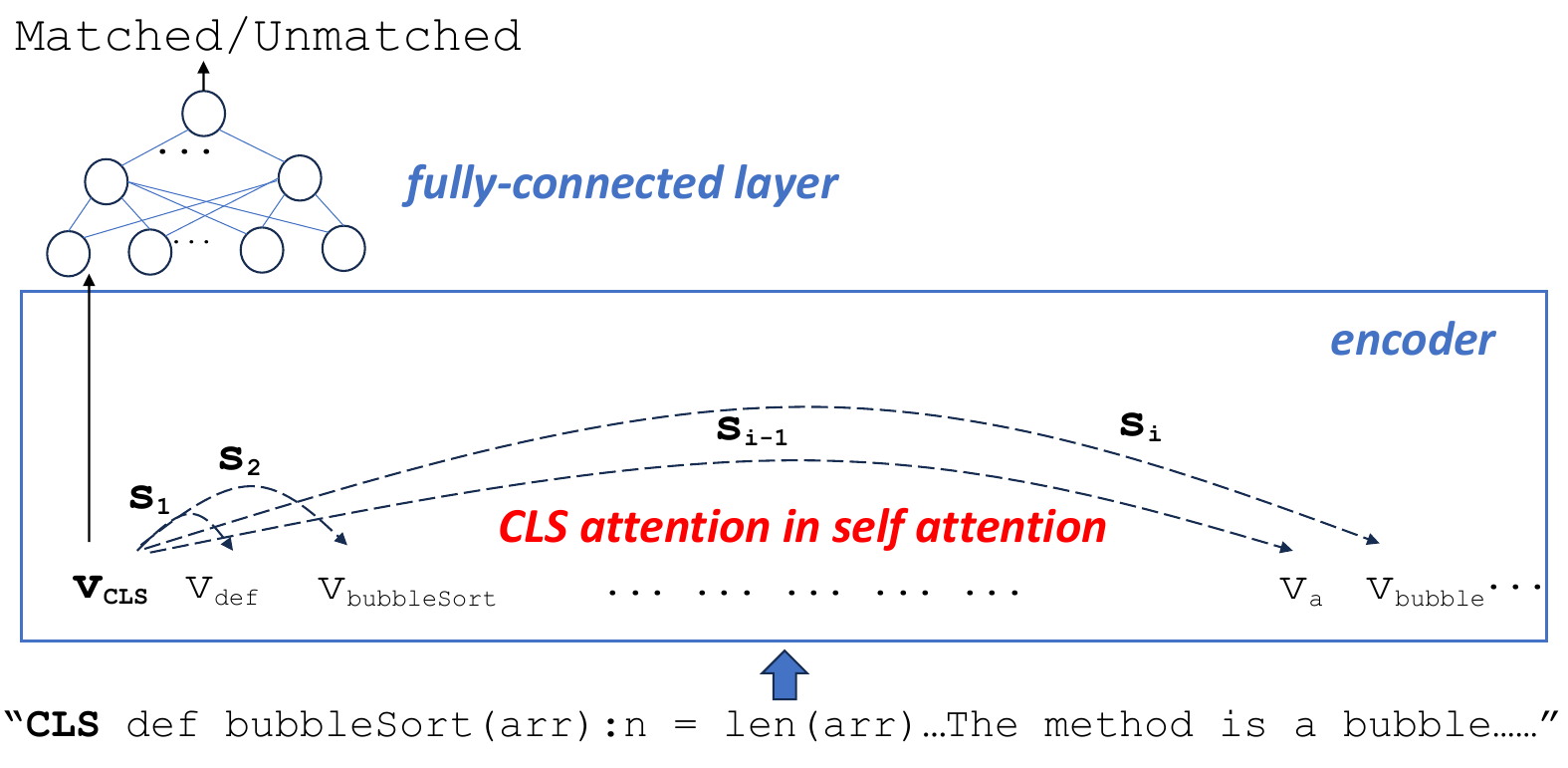}
   \caption{Example of the CLS-attention. It is the self-attention of the `CLS' token that is calculated by the weighted sum of attention over all tokens in the input sequence to derive the representation of the CLS token. 
   The input is the concatenation of a code snippet and a description, the output is `matched' or `unmatched', $v$ is the vector of each token and \color{black}{$s$s are the attention scores of  `CLS' token on other tokens based on Equation~\ref{equ:cls-attention}.}}
   \label{fig:cls-att} 
\end{subfigure}
\hspace{3pt}
\begin{subfigure}[b]{0.48\textwidth}
   \centering
   \includegraphics[width=\textwidth]{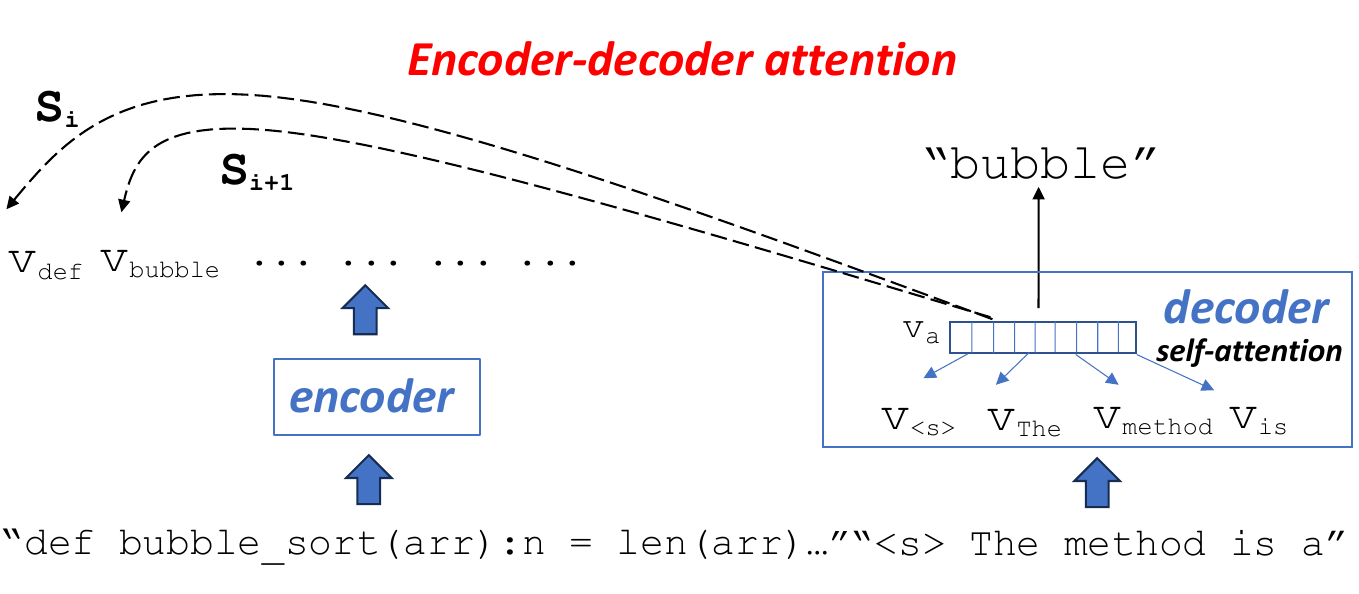}
   \caption{Example of Encoder-Decoder attention. The attention is computed by paying attention to the encoder's output while also maintaining self-attention within the decoder layers to generate context-aware representations for the target sequence. 
   The vector `$v_a$' first performs self-attention with the generated vectors $v_{<s>}$, $v_{The}$, $v_{method}$, $v_{is}$ then absorbs the vectors of input tokens through encoder-decoder's weights to generate next token "bubble".}
   \label{fig:en-de-att}
\end{subfigure}
\vspace{-8pt}
\caption{CLS and Encoder-Decoder attention scores based on the example in Fig.~\ref{fig:motiv-example}.}
\label{fig:attns}
\end{figure*}

For code summarization like task, DietCode still uses the attention scores of this encoder to signify the importance of code tokens. However, it overlooks a crucial aspect: {\em the encoder-decoder attention}. Fig.~\ref{fig:en-de-att} illustrates this encoder-decoder attention mechanism during the generation for the code in Fig.~\ref{fig:motiv-example}. As shown, the vector $v_{bubble}$ should receive more attention from the decoder, and the product of $v_{bubble}$ and normalized {\color{black}{attention score}} $s_{i+1}$ is integrated into the final vector $v_a$ to generate the token `bubble' (after the previous tokens have already been generated). Thus, input vectors with higher decoder-attention scores typically hold greater importance for sequence-to-sequence tasks and can effectively discern the tokens' significance. However, DietCode uses only self-attention scores and discarded these encoder-decoder attention scores.

\subsection{An Example of CLS and encoder-decoder attentions}
\label{subsec:clsendeattion}

Let us use an example to illustrate the problem and to motivate our work (Sections~\ref{sec:empirical-study} and~\ref{sec:algorithm}).


The state-of-the-art \toolbase~\cite{dietcode-fse22} uses the accumulated attention scores to indicate the importance of each input token for downstream tasks. For classification tasks, \toolbase uses a pre-trained encoder (e.g., CodeBERT or CodeT5) in conjunction with a fully-connected layer for code search. For generation tasks, e.g., code summarization, it uses a sequence-to-sequence structure that combines either the encoder of CodeBERT or CodeT5 with a Transformer decoder or CodeT5 decoder. For both tasks, the calculation of the importance of a token is only based on the self-attention scores of the encoder in \toolbase.  
Specifically, the accumulated self attention score of an input token is calculated in Equation~\ref{equ:self-attention}.
Given an input sequence $X = [x_1$, $x_2$, $\ldots$, $x_n]$ where $x_i$ is a $d$-dimensional vector. First, Query ($Q= [q_1, q_2, \ldots, q_n]$), Key ($K= [k_1, k_2, \ldots, k_n]$), and Value ($V$= $[v_1$, $v_2$, $\ldots$, $v_n]$) vectors for a token are generated via three linear mappings. These mappings are implemented by the weight matrices $W^Q$, $W^K$, and $W^V$, which are learnable parameters.
For any token $x_i$, its key vector is dot-producted with every query vector in the sequence, yielding an accumulated attention score that is scaled down by the square root of $d$ as shown in Equation~(\ref{equ:self-attention}).
\begin{equation}
   s_{i} = \frac{\sum_{j=1}^{n} q_j \cdot k_i}{\sqrt{d}}.
\label{equ:self-attention}
\end{equation}
Each accumulated attention score $s_i$ measures how the corresponding token gains attention from other tokens in the input sequence. 
The example of the accumulated self-attention based for the example in Fig.~\ref{fig:motiv-example} is shown in Fig.~\ref{fig:self-attention}. 

\begin{figure*}[t]
\centering
 \lstset{
		numbers=left,
		numberstyle= \tiny,
		keywordstyle= \color{blue!70},
		commentstyle= \color{red!50!green!50!blue!50},
		frame=shadowbox,
		rulesepcolor= \color{red!20!green!20!blue!20} ,
		xleftmargin=1.5em,xrightmargin=0em, aboveskip=1em,
		framexleftmargin=1.5em,
                numbersep= 5pt,
		language=Python,
    basicstyle=\scriptsize\ttfamily,
    numberstyle=\scriptsize\ttfamily,
    emphstyle=\bfseries,
                moredelim=**[is][\color{red}]{@}{@},
		escapeinside= {(*@}{@*)}
	}
\begin{minipage}{\textwidth}
\begin{framed}
{\em The code of bubble sort:}
\begin{lstlisting}
def bubbleSort(arr):
    n = len(arr)
    for i in range(n - 1):
        for j in range(n - 1 - i):
            if arr[j] > arr[j + 1]:
                arr[j], arr[j + 1] = arr[j + 1], arr[j]
\end{lstlisting}
{\em The description:}
    {\em "The method is a basic implementation of the bubble sort algorithm in Python. Bubble sort works by repeatedly swapping adjacent elements if they are in the wrong order until the entire array is sorted."}
\end{framed}
\caption{An Example of Bubble-Sort Code} 
\label{fig:motiv-example}
\end{minipage}
\begin{minipage}{.55\textwidth}
  \centering
  \includegraphics[width=3.1in]{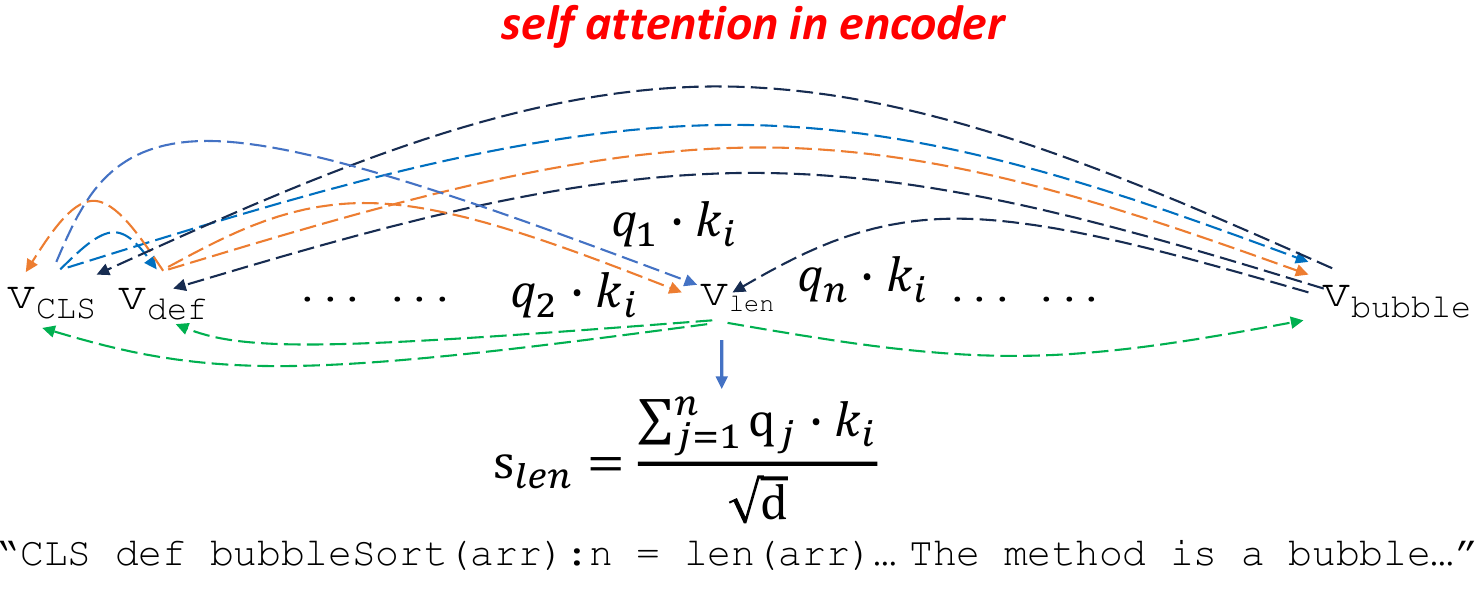}
  \caption{Example of {\color{black}{accumulated}} self attention in $\toolbase$ on Fig.~\ref{fig:motiv-example}. {\color{black}{Dotted lines with different colors represent the self-attentions of different tokens. $s_{len}$ is the accumulated self-attention score of the token {\em `len'}.}}}
  \label{fig:self-attention-score}
\end{minipage}
\caption{An Example of {\color{black}{Accumulated}} Self-Attention in $\toolbase$ on Fig.~\ref{fig:motiv-example}.}
  \label{fig:self-attention}
\end{figure*}

For the classification tasks, e.g., \underline{code search}, {\em only the `CLS' token is sent into the fully-connected layer for label prediction} as shown in Fig.~\ref{fig:attns}a. Thus, {\bf the tokens to which the `CLS' token pays attention could play more crucial roles in the classification task than other tokens}. 
For the CLS attention score $s_i$ for corresponding token, it can be calculated in Equation~\ref{equ:cls-attention}:
\begin{equation}
       s_i = \frac{q_{cls} \cdot k_i}{\sqrt{d}}, (1 \le i \le n)
\label{equ:cls-attention}
\end{equation}

In code search, the model assesses the correspondence between the code and its accompanying description. Via pre-training or fine-tuning, the model acquires an understanding of the interrelation among tokens in bi-modal data via self-attention mechanisms. For instance, the code token {\em "def"} exhibits considerable attention toward tokens like {\em "the", "method", "is", "a"}, and {\em "basic"} within the description. Conversely, the token {\em "method"} in the description concentrates its attention scores on the tokens within the method signature, such as {\em "def"}, {\em "bubble"}, and {\em "method"}. Meanwhile, tokens such as \code{+}, \code{=}, and \code{if} lack clear respective tokens in the text. Thus, the removal of such tokens may not notably impact the matching outcomes, as the models heavily rely on mapping information.


Figures~\ref{fig:cls-en-de}a) and b) represent CLS attention scores on the tokens in the code and the description, respectively. The token <s> serves as the `CLS' token. The deeper the color, the higher the value of the attention score. As seen in Fig.~\ref{fig:cls-en-de}a), the CLS attention (the self attention of the token `CLS') assigns greater importance to the tokens with more bi-modal mappings to form its vector representation. The {\bf `CLS' token allocates significant attention to most tokens in the textual description, while emphasizing on the code tokens primarily within the method signature}. 


\begin{figure*}
    \centering
\includegraphics[width=0.95\textwidth]{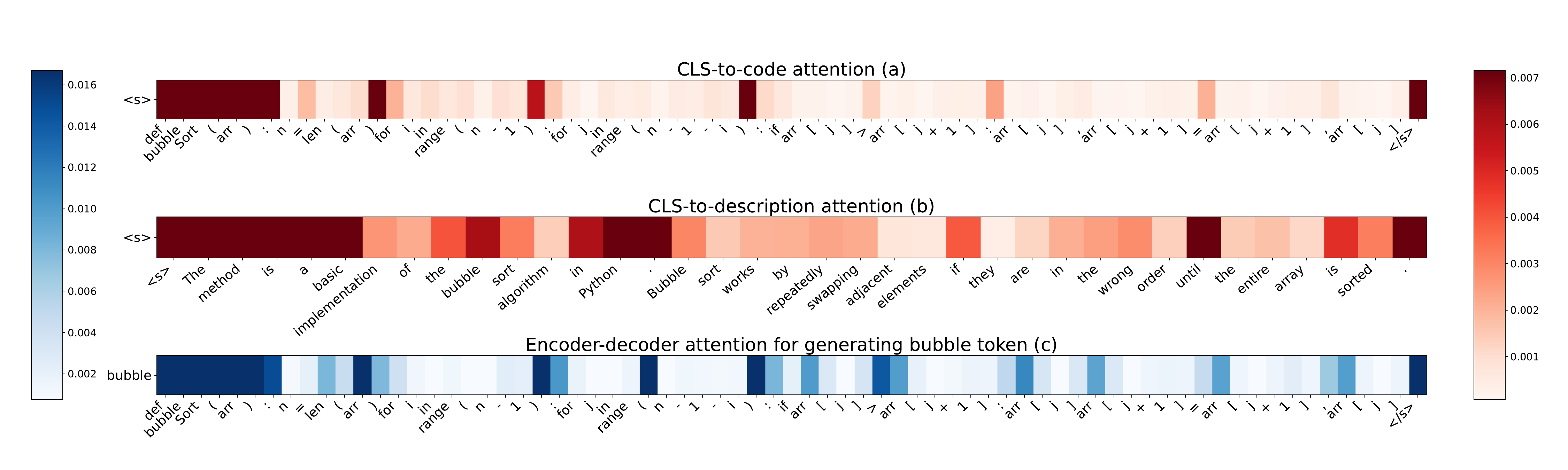}
        \vspace{-14pt}
    \caption{`CLS' and encoder-decoder attentions on the input of the example of bubble sort. (<s> is the CLS token. Deeper color shows higher attention score. Fig.~\ref{fig:cls-en-de}a) and b) represent the CLS attentions on the code tokens and the description, and Fig.~\ref{fig:cls-en-de}c) shows the encoder-decoder attention for the generation of the "bubble"~token.)
}
\label{fig:cls-en-de} 
\end{figure*}

Similarly, for \underline{code summarization}, the bi-modal mappings are apparent. In Fig.~\ref{fig:cls-en-de}c), when the next generated token is the token {\em "bubble"} (part of the method name) in the description, the encoder-decoder attention should emphasize on the tokens of method signature in the code. Fig.~\ref{fig:cls-en-de}c) shows the encoder-decoder attention. As seen, {\bf all tokens in the method signature are paid much attention than others}. 
More specific, with the current query vector $q_t$ at position $t$ in the decoder, the Encoder-Decoder score for each input token is calculated via Equation (\ref{equ:en-de-attention}):
\begin{equation}
   s_i = \frac{q_t \cdot k_i}{\sqrt{d}}, (1 \le i \le n)
\label{equ:en-de-attention}
\end{equation}
Thus, maintaining the mappings between bi-modal data is crucial for bi-modal tasks, which can be accomplished using Encoder-Decoder attention. However, 
\toolbase leverages the attention across all tokens, without giving emphasis to bi-modal mappings.

\subsection{Detailed Results and Analysis of 3 Research Questions in Preliminary Empirical Study}

\subsubsection{\bf (RQ-1) What important tokens do CLS attentions emphasize on?}

In Table~\ref{tab:cls-var-appendix}, the five leftmost columns show the categories, the maximum of CLS attention scores, the minimum of CLS attention scores, the averages and variances of the global attention scores of tokens for each category. The global attention scores are used to calculate the average attention score of each token within a statement category. The last two columns
of Table~\ref{tab:cls-var-appendix} display our proposed averages of 
category-local attention scores, and the average variances of 
CLS attention scores of tokens grouped by statement categories.
There are 21 categories used by \toolbase, collectively representing over 95\% of the statements in the dataset. In {\tool}, we use these categories as contexts for tokens. They offer relatively straightforward setups for bi-modal mappings.

Analyzing the columns reveals that the category-local attention average of tokens within method signatures significantly surpasses that of others, even exceeding the second largest class (‘return’ statements) by a factor of eight. The \code{Return} class typically comprises the names of returned variables, which often convey crucial functional information. Logging and Annotation classes rank third and fourth, respectively, as they frequently encompass function-related data. Variable Declaration and Function Invocation are next, 
predominantly due to inclusion of variables or callee names.

Conversely, the \code{Break}, \code{Case}, and \code{Continue} exhibit the lowest category-local attention average. This is attributed to the fact that \code{Break} and \code{Continue} statements have keyword information, while \code{Case} statements, though potentially containing conditions, may be too detailed to establish meaningful bi-modal mappings with the description. Moreover, the average variance is directly proportional to the average attention scores, indicating that higher attention scores correspond to increased~variance.
 
Comparing our proposed averages (last two columns) with the global averages (the 4$^{th}$ and 5$^{th}$ columns)
in Table~\ref{tab:cls-var-appendix}, except Method signature, the averages of attention scores/variances for the 
context-aware, category-local attentions as we propose
are reduced from 0.55/5 times to 3.3/844 times in comparison to the global averages within each statement category. The variance of Method Signature increases 3.3 times when the average of attention scores goes up more than 4 times.
Thus, {\bf the categories of the statements successfully reflect the context of each token}.

\begin{table*}
\centering
\caption{{\color{black}{(RQ-1)}} \color{black}{Statistics of CLS attention scores on 0.9M training dataset. (Max/Min: the max/min of CLS attention scores in each category; 
 Global/Global\_variance: the average/variance of global attention scores for each category; Category-local/Local\_variance: the averages/variance of category-local attention scores.)}}
\vspace{-8pt}
\tabcolsep 3.5pt
\begin{tabular}{ccccccc}
\hline
Category & Max & Min & {\bf Global} & {\bf Global\_variance} & {\bf \color{black}{Category-local}} & {\bf \color{black}{Local\_variance}} \\ 
\hline
Annotation&13.64774&0.00369&0.22834&1.09668&0.14712&0.25699\\
Arithmetic&3.83190&0.00269&0.23538&1.15031&0.05800&0.00915\\
Variable Declaration&10.49360&0.00353&0.24737&1.18534&0.10431&0.10301\\
Function Invocation&5.25585&0.00349&0.25464&1.23739&0.10638&0.09292\\
Return&6.59062&0.00353&0.27754&1.33069&0.20165&0.17654\\
Switch&4.36353&0.00352&0.23530&1.02716&0.07194&0.00936\\
Break&0.83018&0.00536&0.23083&1.02695&0.04734&0.00253\\
Setter&2.85442&0.00196&0.24565&1.16452&0.07033&0.04076\\
Synchronized&1.02363&0.00526&0.23753&1.00888&0.08507&0.01309\\
Try&4.33814&0.00371&0.24224&1.05390&0.08925&0.02697\\
Catch&2.07879&0.00385&0.24606&0.90369&0.05224&0.00641\\
Method Signature&29.17083&0.00353&0.33731&1.56490&1.74525&6.69616\\
Finally&0.53785&0.00439&0.22154&1.41311&0.09047&0.00939\\
Getter&7.59034&0.00373&0.24690&1.18191&0.06407&0.03392\\
Throw&5.20869&0.00346&0.23824&1.12604&0.08951&0.02551\\
Case&2.70762&0.00338&0.23824&1.15401&0.03953&0.00673\\
While&2.57405&0.00311&0.22953&1.06577&0.05870&0.01045\\
Continue&0.44534&0.00662&0.25256&1.35214&0.04609&0.00160\\
If Condition&4.03009&0.00262&0.24831&1.21111&0.08341&0.04783\\
For&22.29719&0.00348&0.24910&1.11164&0.07938&0.02098\\
Logging&4.90106&0.00278&0.24053&1.15288&0.14864&0.11938\\
\hline
\end{tabular}
\label{tab:cls-var-appendix}
\end{table*}

\subsubsection{\bf (RQ-2) What important tokens do encoder-decoder attentions emphasize about?}

\begin{table*}
\centering
\caption{{\color{black}{(RQ-2)}} \color{black}{Statistics of encoder-decoder attention scores based on 0.16M training dataset. (Max/Min: the maximum/minimum of encoder-decoder attention scores in each category; Global/Global\_variance: the average/variance of the global attention scores of tokens for each category; Category-local/Local\_variance: the averages/variance of category-local attention scores.)}}
\vspace{-4pt}
\tabcolsep 3.5pt
\begin{tabular}{ccccccc}
\hline
Category & Max & Min & {\bf Global} & {\bf Global\_variance} & {\bf \color{black}{Category-local}} & {\bf \color{black}{Local\_variance}} \\ 
\hline
Annotation&7.94088&0.31593&2.61080&13.75766&1.54646&0.08766\\
Arithmetic&37.44108&0.06535&2.68657&15.41490&2.30496&2.52065\\
Variable Declaration&65.53831&0.08960&2.86623&15.63358&2.69239&7.96826\\
Function Invocation&63.96667&0.00918&2.86456&15.94000&2.80177&8.54368\\
Return&55.23188&0.09667&3.08082&17.61045&4.75692&16.01764\\
Switch&30.02598&0.07416&2.70932&16.36071&2.40672&2.63213\\
Break&28.02130&0.04057&2.64491&16.43219&2.66594&1.21296\\
Setter&69.05594&0.02932&2.84634&17.25399&2.32736&5.09966\\
Synchronized&78.08508&0.04113&2.84346&17.08405&3.11007&3.03446\\
Try&78.26600&0.02602&2.82429&17.31111&2.45927&2.69108\\
Catch&34.98584&0.07141&3.01365&19.80343&2.44455&4.18487\\
Method Signature&91.68832&0.13817&3.29116&18.21317&5.91461&30.92008\\
Finally&10.49109&0.74180&2.38186&7.77507&2.98907&1.73634\\
Getter&68.48614&0.02599&2.87695&16.58745&2.57884&6.41530\\
Throw&87.67154&0.05743&2.79540&16.04224&3.09842&8.12932\\
Case&23.25289&0.03370&2.74530&16.10519&1.79642&1.54551\\
While&67.68087&0.04015&2.69720&15.52064&2.40500&3.14009\\
Continue&9.85333&0.26809&2.48510&12.64186&1.72747&0.36762\\
If Condition&57.87812&0.05160&2.83536&15.84334&2.49570&5.97373\\
For&60.61899&0.03448&2.90877&17.21421&2.99445&6.88533\\
Logging&65.63157&0.03885&2.77210&15.53356&2.89339&8.41999\\
\hline
\end{tabular}
\label{tab:en-de-var-appendix}
\end{table*}

Table~\ref{tab:en-de-var-appendix} shows the average of the Encoder-Decoder attention scores of tokens based on statement types, called {\bf Encoder-Decoder category-local attention average}.
Unlike CLS attention, each token in the input can have multiple Encoder-Decoder attention scores, i.e., for each generated token, the decoder calculates an attention score for each token in the input. For example, Fig.~\ref{fig:en-de-att} shows the Encoder-Decoder attention scores for the input at the generation of the token "bubble". Thus, the largest attention score is chosen as the attention score. 

As depicted in Table~\ref{tab:en-de-var-appendix}, the categories with the highest and lowest importance remain consistent as `Method signature', \code{Return}, \code{Continue}, and \code{Case}, respectively. However, the disparity between them has diminished. Certain categories, e.g., \code{Synchronized}, \code{For}, and \code{Throw}, have surpassed `Logging', `Variable Declaration', and `Function Invocation', securing the 3$^{rd}$, 4$^{th}$, and 5$^{th}$ places in importance. This shift indicates a redistribution of importance across categories.
The Encoder-Decoder attention scores are generated in conjunction with the description. In the instances where the description contains intricate function details, these tokens garner high attention scores, facilitating the establishment of bi-modal mappings. For code search, the significance of details within the code (e.g., \code{Throw} statements) is lower compared to the broader functional description (e.g., `Method signature').

Similar to Table~\ref{tab:cls-var-appendix}, except `Method' signature, comparing the 4$^{th}$ and 5$^{th}$ columns with the last two columns, our proposed averages of attention scores/variances for context-aware, category-local attentions are much reduced from 0.1 to 156 times compared to the global averages within each category. However, the change from the global attention average to the category-local average is insignificant. This could be due to 1) our selection of the largest attention score as the representative, and 2) the difference in the Encoder-Decoder attention scores is not substantial across categories.

\subsubsection{\bf (RQ-3) Do the averages of self-attention scores reflect
the CLS attentions and the Encoder-Decoder attentions?}
\label{subsubsec:diff}
Our answer is `No'.
{\em The accumulated attention scores from the self-attention (as used in $\toolbase$) is for pre-training and cannot reflect and substitute for those from the CLS and Encoder-Decoder attentions. i.e., the self-attention is used for pre-trained tasks and vectored general representations, not directly for downstream tasks}. For elaboration, these attention schemes are for different tasks. The self attention is for pre-training tasks, while CLS attention is for fine-tuning downstream discriminative tasks, and the Encoder-Decoder attention is for downstream sequence-to-sequence generation tasks. In fact, the encoders of CodeBERT and CodeT5 have been trained in multiple pre-trained tasks. CodeBERT is pre-trained with two objectives: Masked Language Model (MLM, bimodal data) and Replaced Token Detection (RTD, unimodal data). In MLM, the model is trained to predict the identity of tokens that have been randomly masked in the input sequence. In RTD, the model is given an input sequence where some tokens have been replaced with incorrect ones. The model task is to predict which tokens are the original ones and which have been replaced. These objectives enable the model to capture the long dependencies between tokens in bimodal sequence (MLM) and unimodel (RTD) to obtain general representations via the self attention.
Thus, the averages of self-attention scores cannot replace the CLS and Encoder-Decoder attentions.
The latter attentions are directly applied to downstream tasks~(Fig.~\ref{fig:attns}).


To illustrate our above answer, Figures~\ref{fig:heatmaps}a) and~\ref{fig:heatmaps}b) show the heatmaps of the accumulated self-attention (in \toolbase) and the Encoder-Decoder attention for the bubblesort example. Each row is one statement without indentation (except that lines 5-6 contain one statement since it is too long to show). The darker color means higher attention scores, i.e., the tokens are more important.

\begin{figure}
    \centering
     \begin{subfigure}[b]{0.48\textwidth}
         \centering
         \includegraphics[width=1.2\linewidth]{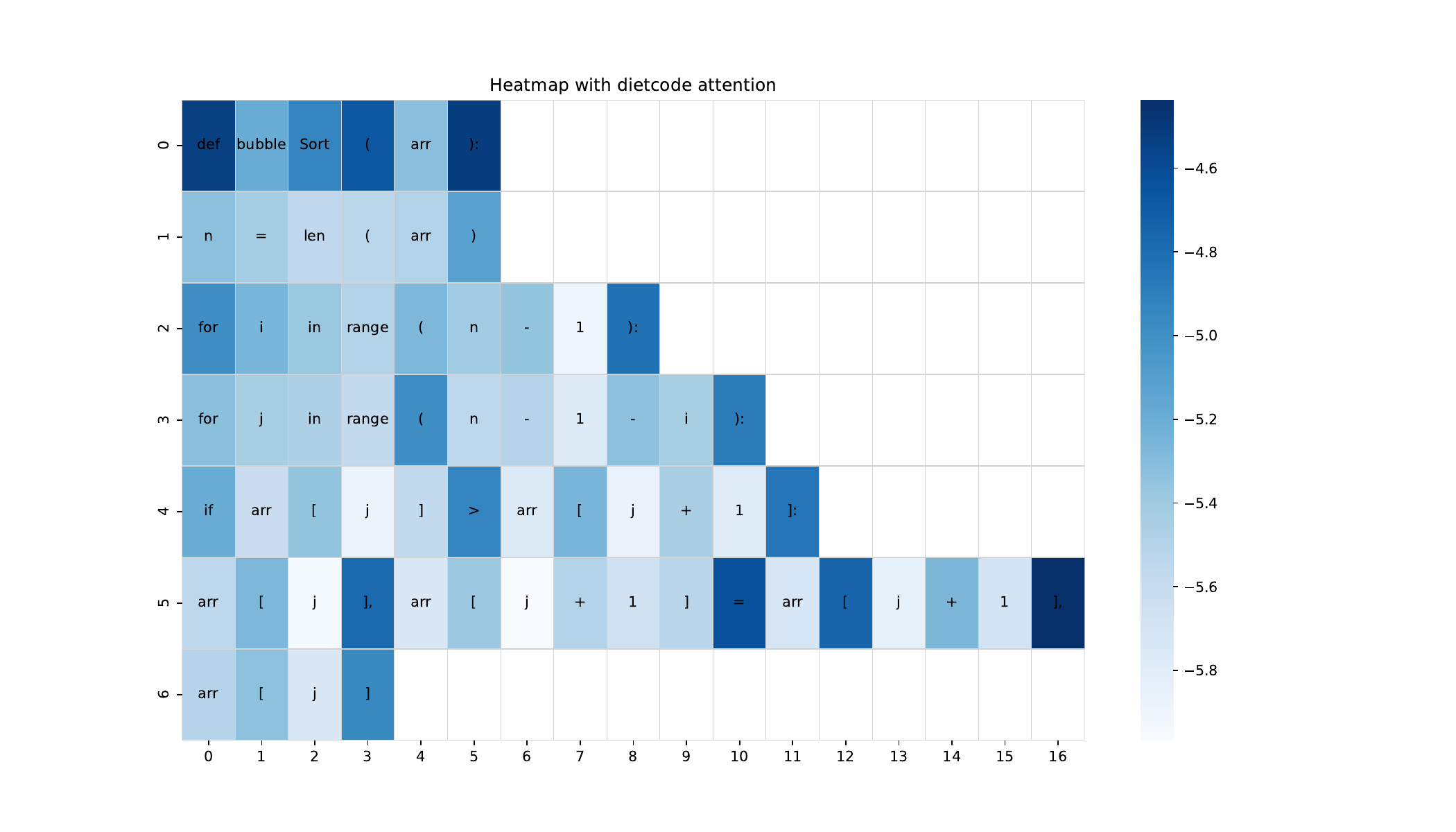}
         \vspace{-12pt}%
         \label{fig:Heatmap_dietcode}
     \end{subfigure}
     \vspace{3pt}
     \hspace{3pt}
     \begin{subfigure}[b]{0.48\textwidth}
         \centering
         \includegraphics[width=1.2\linewidth]{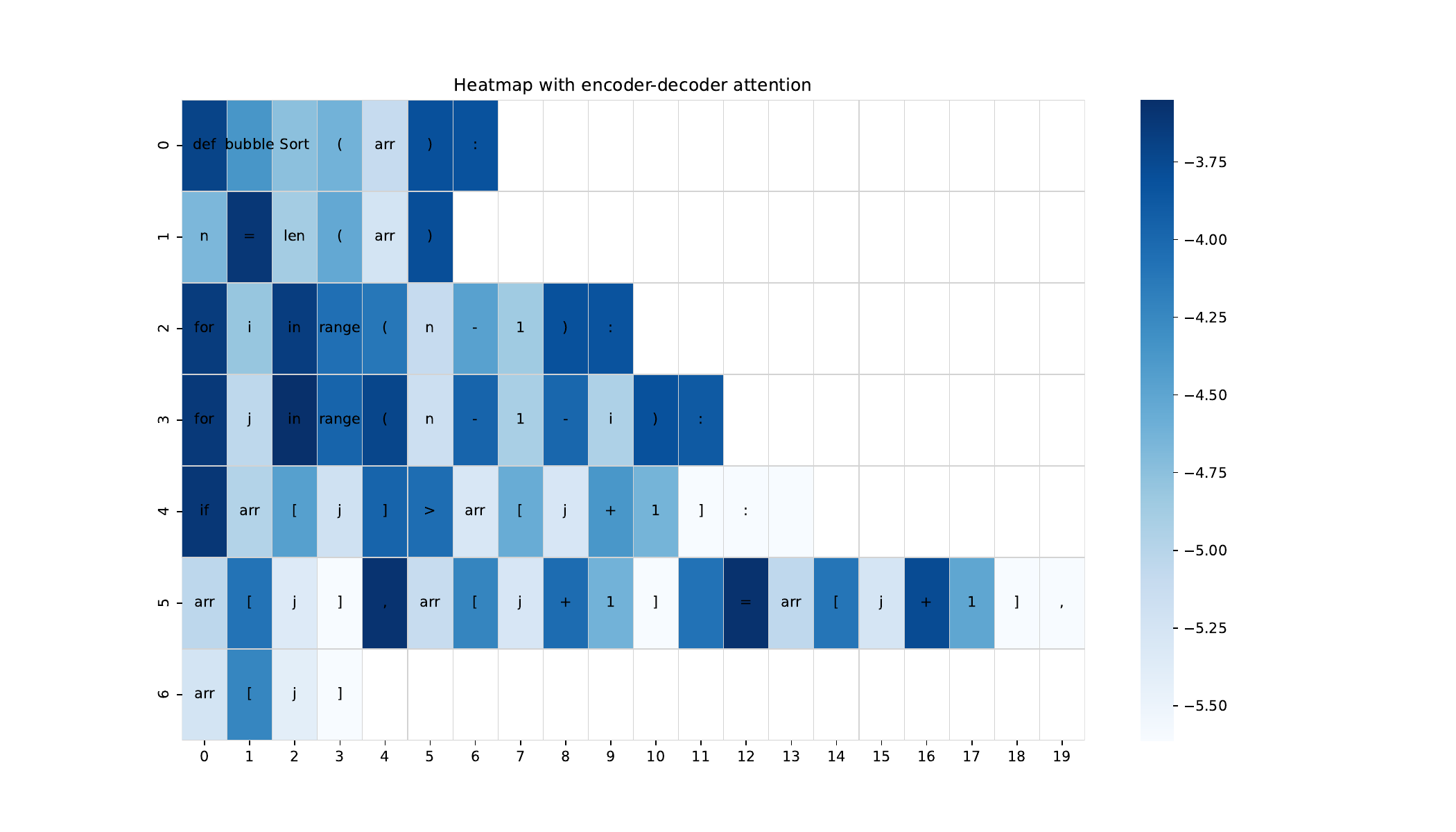}
         \vspace{-12pt}
         \label{fig:Heatmap_en}
     \end{subfigure}
     \vspace{-12pt}
        \caption{Bubble-sort: Heatmaps of (a) accumulated self-attention scores, (b) Encoder-Decoder attention scores}
        \label{fig:heatmaps}
\end{figure}


For comparison, the colors of the cells in Fig.~\ref{fig:heatmaps}a) (accumulated self-attention used by \toolbase) have the shades in-between the ones of Figure~\ref{fig:cls-en-de}a) for `CLS' attention and Fig.~\ref{fig:heatmaps}b) for Encoder-Decoder attention.
Specifically, for both method signature and method body, both contain tokens of higher importance as in Fig.~\ref{fig:heatmaps}b). The importance of tokens in the method signature is slightly higher than that of tokens in the method body in Fig.~\ref{fig:cls-en-de}a).
The above situation mainly stems from MLM and RTD tasks needing to do token-level prediction in code, thus requiring to pay attention on tokens in the method body. Keywords and separators can have high attention weights as they play important roles in predicting tokens~\cite{dietcode-fse22}. In addition, since tokens in the method signature could contain more important `guidance' for prediction, their attention scores could be larger.



In Fig.~\ref{fig:heatmaps}b), at each text token generation iteration, every code token has one attention score, and here the largest attention score in all iterations is used since the largest score represents the most important degree that the token has played in the generation process. As seen, the method body and signature are almost equally important, and import tokens are more evenly distributed for this code snippet. Namely, different generated tokens pay attention to different code tokens. This is largely different from the CLS attention distribution in code search, as shown in Fig.~\ref{fig:cls-en-de}a), which emphasizes on the tokens mainly in the method signature.





\subsection{Dataset Statistics}
Table~\ref{tab:statistics_code} shows the detailed dataset statistics.
\begin{table}[t]
    \centering
    \small
    \caption{\underline{Dataset Statistics} (Total: \# code snippets, Avg/Max/Min: the average/max/min \# of tokens in a code snippet or description, tr: training, val: validation, test: test dataset, Search: code search, Sum: summarization).}
    \vspace{-6pt}
    \setlength{\tabcolsep}{1.5pt}
    \resizebox{\columnwidth}{!}{%
    \begin{tabular}{l|cccc|ccc}
        \hline
        \multirow{2}{*}{\textbf{Dataset}}& \multicolumn{4}{c|}{\textbf{Code Snippet}} & \multicolumn{3}{c}{\textbf{Code Description}} \\
        \cline{2-8}
             & {\bf Total} & {\bf Avg} & {\bf Max}  & {\bf Min}  & \textbf{Avg}  & \textbf{Max}  & \textbf{Min} \\
        \hline
        Search\_{tr} & 908,886& 112.67 & 68,278 & 20 & 19.2 & 3439 & 1 \\
        \hline
        Search\_{val} & 30,655& 95.57 & 3,092 & 21 & 19.05 & 521 & 1 \\
        \hline
        Search\_{test} & 26,909& 113.42 & 5,542 & 20 & 20.22 & 709& 1 \\
        \hline
        Sum\_{tr} & 164,923& 100.99 & 512 & 17 & 13.25 & 175 & 3 \\
        \hline
        Sum\_{val} & 5,183 & 90.79 & 501 & 18 & 13.39 & 147 & 3 \\
        \hline
        Sum\_{test} & 10,955& 100.06 & 512 & 20  & 12.71 & 111 & 3 \\
        \hline
    \end{tabular}
    }
    \label{tab:statistics_code}
\end{table}

\subsection{Replacement Study}

\begin{table}[t]
  \centering
  \small
  \caption{Results of \textbf{\underline{\tool}} with {\toolbase}'s removal algorithm (10\%-50\% removal for each snippet, BLEU: BLEU-4 values, R-B:Reduced BLEU-4, R-M: Reduced MRR)}
  \label{tab:DietCodeRemoveTokenAlgorithmResults}
  \vspace{-6pt}
  \setlength{\tabcolsep}{1pt} 
  \renewcommand{\arraystretch}{0.9} 
  \resizebox{\columnwidth}{!}{%
  \begin{tabular}{c|cc|cc|cc|cc}
    \hline 
    \multicolumn{1}{c}{} & \multicolumn{4}{c|}{\textbf{Code Search}} & \multicolumn{4}{c}{\textbf{Code Summarization}} \\
    \hline 
    \multicolumn{1}{c|}{\textbf{Ratio}} & \multicolumn{2}{c|}{\textbf{CodeBERT}} & \multicolumn{2}{c|}{\textbf{CodeT5}} & \multicolumn{2}{c|}{\textbf{CodeBERT}} & \multicolumn{2}{c}{\textbf{CodeT5}} \\
    & MRR & R-M & MRR & R-M & BLUE & R-B & BLUE & R-B \\
    \hline 
    Base  & 0.726& --- & 0.747 & --- & 18.25& --- & 20.55& --- \\
    10\%  & 0.701& 3.44\%\color{green}{↓}& 0.723& 3.21\%\color{green}{↓}& 17.40& 4.66\%\color{green}{↓}& 18.53& 9.83\%\color{green}{↓}\\
    20\%  &  0.703& 3.17\%\color{green}{↓}&  0.717& 4.01\%\color{green}{↓}&  17.34& 4.97\%\color{green}{↓}& 18.18& 11.53\%\color{green}{↓}\\
    30\%  & 0.702& 3.31\%\color{green}{↓}& 0.712& 4.69\%\color{green}{↓}& 17.31& 5.15\%\color{green}{↓}& 18.08& 12.02\%\color{green}{↓}\\
    40\%  &  0.696& 4.13\%\color{green}{↓}&  0.713& 4.55\%\color{green}{↓}&  17.09& 6.35\%\color{green}{↓}& 18.03& 12.26\%\color{green}{↓}\\
    50\%  & 0.682& 6.06\%\color{green}{↓}& 0.695& 6.96\%\color{green}{↓}& 16.73& 8.32\%\color{green}{↓}& 17.63& 14.20\%\color{green}{↓}\\
    \hline 
  \end{tabular}
  }
  \label{tab:replacement}
\end{table}

{\color{custom-blue}{\tool mainly consists of two aspects: token weights and a token removal algorithm (Algorithm~\ref{alg:leancode}).
Here, we aim to determine how each of them contributes to \tool's performance. Our procedure is to replace \tool's removal algorithm in {\tool} with \toolbase's. 
Then, comparing the replacement results with the \toolbase's results, the performance drop will reflect the contribute of token weights for \tool' performance due to the same removal algorithm. Similarly, the difference of results of \tool and the replacement method can show the contribute of \tool's removal algorithm since both share the same token weights.}}

{\color{custom-blue}{Table~\ref{tab:replacement} shows the replacement results. Compared to \toolbase'results in Tables~\ref{tab:CodeSearch} and~\ref{tab:CodeSummarization}, we can see that the performance drops of \toolbase reach up to 37.1\% for code search and 19.05\% for code summarization. The performance drop conversely demonstrates that more accurate token weights significantly enhanced \tool's performance.
On the other side, compared to \tool' results, we find that the maximal performance drops can be 3.7\% for code search and 9.9\% for code summarization. It demonstrates that the token-level removal algorithm (Algorithm~\ref{alg:leancode}) is much better than that of \toolbase and significantly contributes to {\tool}'s performance.}}

\end{document}